\documentclass[useAMS,usenatbib]{mn2e}

\usepackage{graphicx}
\usepackage{times}
\usepackage{color}
\newcommand{\kms}{\mbox{$\mathrm{km~s^{-1}}$}}
\newcommand{\mum}{\mbox{$\mathrm{\mu m}$}}
\newcommand{\Msun}{\mbox{M$_\odot$}}
\newcommand{\Lsun}{\mbox{L$_\odot$}}

\newcommand{\note}[1]{\textcolor{black}{#1}}

\title[NGC\,4438's nuclear region]{Near-IR imaging and spectroscopy of
  the nuclear region of the disturbed Virgo cluster spiral NGC\,4438
  \thanks{Based on VLT service mode observations (Programme 69.B-0411)
    and TIMMI2 (run 68.D-0432) gathered at the European Southern
    Observatory, Chile.}}  \author[Perez et al.]{Sebastian Perez
  M.$^{1,2}$\thanks{E-mail: s.perez2@physics.ox.ac.uk}, Simon
  Casassus$^{1,5}$, Juan R. Cort\'es$^{1,3}$ and Jeffrey D. P.
  Kenney$^{4}$ \\ $^{1}$Departamento de Astronom\'ia, Universidad de
  Chile, Casilla 36-D, Santiago, Chile\\ $^{2}$University of Oxford,
  Department of Physics, Keble Road, Oxford, OX1 3RH, U.K \\ $^{3}$
  Joint ALMA Observatory, Casilla El Golf 16-10, Las Condes, Santiago,
  Chile\\ $^{4}$Department of Astronomy, Yale University, New Haven \\
  $^{5}$LUTH, Observatoire de Paris, CNRS, Universit\'e Paris Diderot,
  5 Place Jules Janssen, 92190 Meudon, France}

\begin{document}
\date{} 

\pagerange{\pageref{firstpage}--\pageref{lastpage}} \pubyear{2009}

\maketitle

\label{firstpage}

\begin{abstract}
  We present near-infrared VLT ISAAC imaging and spectroscopy of the
  peculiar Virgo galaxy NGC\,4438, whose nucleus has been classified
  as a LINER. The data are supplemented by mid-infrared imaging, and
  compared to previous WFPC2 \textit{HST} broadband images. Images and
  position-velocity maps of the [Fe\,{\sc ii}] and H$_2$ line
  emissions are presented and compared with the distribution of the
  optical narrow-line region and radio features. Our results show that
  shocks (possibly driven by a radio jet) contribute to an important
  fraction of the excitation of [Fe\,{\sc ii}], while X-ray heating
  from a central AGN may be responsible for the H$_2$ excitation. We
  address the question whether the outflow has an AGN or a starburst
  origin by providing new estimates of the central star formation rate
  and the kinetic energy associated with the gas. By fitting a
  S\'ersic bulge, an exponential disc and a compact nuclear source to
  the light distribution, we decomposed NGC\,4438's light distribution
  and found an unresolved nuclear source at 0.8$\arcsec$ resolution
  with $M_K = -18.7$ and $J-H = 0.69$. Our measured bulge velocity
  dispersion, 142~\kms, together with the standard $\mathcal{M}_{\rm
    bh}-\sigma$ relation, suggests a central black hole mass of
  $\log(\mathcal{M}_{\rm bh}/\Msun) \sim 7.0$. The stellar kinematics
  measured from the near-infrared CO lines shows a strong peak in the
  velocity dispersion of $\sigma_0\sim$178~\kms\ in the central
  0.5\arcsec, which is possible kinematic evidence of a central black
  hole. We calculated a general expression for the integrated S\'ersic
  profile flux density in elliptical geometry, including the case of
  `disky' isophotes.
\end{abstract}


\section{Introduction}

NGC\,4438 is a large peculiar spiral galaxy, with a disturbed stellar
disc and an even more heavily disturbed interstellar medium
(ISM). Located near the centre of the Virgo cluster, NGC\,4438 has
undergone a violent collision with the nearby giant elliptical M86
\citep{ken08}, and may also be experiencing ongoing ram pressure
stripping due to an interaction with the Virgo intracluster medium
\citep{vol09}. NGC\,4438's nucleus has been classified as a
particularly interesting LINER \citep[low-ionisation nuclear
emission-line region,][]{heck80}. The spectra of LINERs are
characterized by the presence of emission lines from atomic species of
low ionisation state. By definition, LINERs are galaxies which host
nuclei with emission-line ratios that satisfy the following criteria:
[O~{\sc iii}]$\,\lambda 5007$$/\rmn{H}\beta<3$, [O~{\sc i}]$\,\lambda
6300$$/\rmn{H}\alpha>0.05$, and [N~{\sc ii}]$\,\lambda
6583$$/\rmn{H}\alpha>0.5$ \citep{ost05}. Their emission line spectra
are similar to those observed in narrow-line regions (NLRs) of gas in
Seyfert 2 galaxies. However, some LINERs have relatively powerful
central black holes, and tend to have broader emission lines
\citep[e.g., NGC\,1052,][]{ho97}.

The most likely mechanism to explain the excitation in these objects
is photoionization either from an active galactic nucleus (AGN) or
from a strong stellar continuum \citep{ho03}. AGN photoionisation
models fit both the low-ionisation spectra of LINERs and the
high-ionisation spectra of the Seyfert NLRs with similar nuclear
emission, but different nebular conditions \citep[such as the electron
  density and the incident ionising luminosity; see discussion
  by][]{ho03,ost05}.

The nucleus of NGC\,4438 lies at the root of a nuclear bubble,
expanding to the north-west (NW), which has been imaged in
H$\alpha$+[{\sc N\,ii}] emission by \citet{ken02}. In contrast with
other similar systems, it is not clear what is powering this bubble,
since there is neither a strong starburst nor a strong AGN in
NGC\,4438. For example, the galaxy NGC\,2782 posseses a central
starburst (extending over a radius $\sim$200~pc) which provides the
mechanical luminosity that drives its central winds \citep{jog99}. The
Seyfert galaxy NGC\,3079 harbours both a powerful AGN and a
circumnuclear starburst \citep{vei94}, while M82's outflows are driven
only by a nuclear starburst \citep{leh99}. Moreover, there are many
other objects powered by composite systems where a circumnuclear
starburst coexists with a central low-luminosity AGN
\citep{rod05}. \citet{lev03} observed the Seyfert 2 starburst galaxy
NGC\,5135, distinguishing both the AGN (unresolved) and the starburst
(spatially extended over $\sim$200~pc).

\begin{figure}
  \centering\includegraphics[width=\columnwidth]{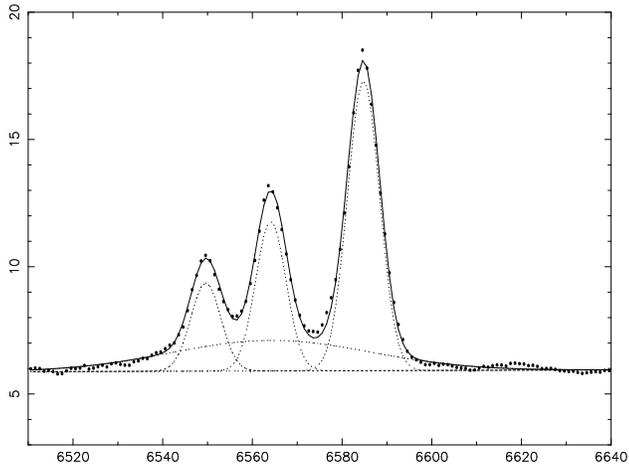}
  \caption{A possible decomposition of H$\alpha$ $\lambda 6562$ and [N
    {\sc ii}] $\lambda \lambda 6548, 6583$ emission, using the minimum
    number of components (a linear baseline, 3 narrow lines and 1
    broad component -- or 13 free parameters). The $x$-axis is
    wavelength in $\mathrm{\AA}$, while the $y$-axis is relative
    intensity. The solid circles are the data points, the solid line
    is the best fit model, and the dotted lines are the contributions
    from each component. Data courtesy of L. Ho.}
  \label{fig:ha_nii}
\end{figure}

Broad H$\alpha$ emission in NGC\,4438 has been tentatively inferred by
\citet{ho97} from the fitting of optical spectra. We have repeated the
fitting of the H$\alpha$ emission line profile on their data but with
a smaller number of Gaussian components than \cite{ho97}, obtaining
similar results. Fig.~\ref{fig:ha_nii} shows that a broad,
FWHM$\sim$2050~\kms, Gaussian component is clearly present in the
emission-line complex. This spectral feature is thought to be
indicative of the presence of active nuclei harboring a broad
emission-line region \citep{ho97}.

Near-IR (near-infrared) light can escape high opacity and dusty
environments more easily than H$\alpha$ photons can, making it
possible to detect heavily obscured line-emitting regions.  Near-IR
emission-lines; such as [Fe\,{\sc ii}] 1.257~$\mu$m and H$_2$
2.122~$\mu$m, are less sensitive to extinction than optical lines such
as H$\alpha$. Near-IR observations can provide better extinction
estimates, and can yield new information on the intensity of a
potentially obscured central AGN or starburst.

Results of a near-IR spectroscopic survey of LINER galaxies carried
out by \citet{lar98}, showed that the [Fe\,{\sc ii}] line
($a^4D_{7/2}-a^6D_{9/2}$ at 1.257~$\mu$m) is the most commonly
detected emission-line in LINERs \citep[see also,][]{rod05}. The H$_2$
line ($1-0\mathrm{S}(1)$ at 2.122~$\mu$m) is also a common feature. On
the other hand, the Pa$\beta$ emission-line (or {\sc H\,i 3--5} at
1.282~$\mu$m), which is the strongest near-IR recombination line
available at low redshift in AGN and starburst galaxies, is only found
in emission in 20\% of the LINER galaxies in \citet{lar98}
sample. Br$\gamma$ (or {\sc H\,i 4--7} at 2.166~$\mu$m) is undetected
in all the galaxies of the survey.

NGC\,4438's combination of peculiar spectral features, complex
morphology and the possibility of studying the nuclear bubble at
different wavelengths led us to obtain new near-IR imaging and
spectroscopic data. In this work, we present the results of ISAAC
imaging and spectroscopy of NGC\,4438, focusing on analysis of line
emission maps and on two-dimensional modelling of the galaxy surface
brightness. We want to study the morphology of NGC\,4438 and whether
there is a nuclear point source embedded in the bulge. In
Section~\ref{obs}, we begin by giving a brief description of the
observations. In Section~\ref{results} we discuss the nuclear line
emission based on both imaging and spectroscopy, together with
extinction estimates. In Section~\ref{sec:CO} we derived the
line-of-sight stellar velocity dispersion from CO bands absorption
features. In Section~\ref{pho} we present the 2D decomposition of the
surface brightness of NGC\,4438. In Section~\ref{dis} we discuss and
compare our findings with previous studies in the X-ray, optical and
radio wavelengths, focusing on the energetics of the nuclear source
and the surface brightness modelling results. Appendix~\ref{dr} gives
a detailed description of the data reduction of the imaging data. In
Appendix~\ref{sbm} we describe the surface brightness model used to
fit the light distribution and how we measure structural parameters.

Throughout this paper we use a distance to NGC\,4438, near the centre
of the Virgo cluster, of 16~Mpc (at which $1\arcsec$ corresponds to a
distance scale of 77.6~pc). All the data reduction and analysis were
carried out using the Perl Data Language
\citep[\texttt{http://pdl.perl.org,}][]{gla97}. The stellar kinematics
was derived using IDL (\texttt{http://www.ittvis.com/idl/}).

\section{Observations}\label{obs}

\subsection{Imaging}

\begin{table}
  \caption{ISAAC imaging observing log and calibration factors.}
  \begin{minipage}{\columnwidth}
    \label{tab:calib}
      \begin{tabular}{@{\extracolsep{\fill}}llcccc}
	\hline
	Date   & Filter & Integration & Zero point & $A_{\lambda}$\footnote{Foreground extinction (see Section~\ref{extin}).} & Seeing\footnote{Effective seeing at the observed wavelength, given by the FWHM of the best-fitted Gaussian profile.} \\
        &  & (sec)  & (mag) & (mag)  & (arcsec)\\ 
        \hline
        \multicolumn{6}{c}{broad-band data}\\
	\hline
	Apr 14 2003 & $J$       & 400 & 24.85 & 0.025 & 0.86\\
	Apr 14 2003 & $H$       & 400 & 24.31 & 0.016 & 0.62\\
	Jan 13 2003 & $K_{\rm s}$& 284 & 24.12 & 0.010 & 0.58\\        
        \hline
        \multicolumn{6}{c}{narrow-band data}\\
        \hline
	Apr 14 2003 & $1.26$  & 600 & 21.88 &  --   & 0.69\\
	Apr 14 2003 & $1.28$  & 600 & 21.69 &  --   & 0.56\\
	Jan 13 2003 & $2.07$  & 800 & 21.62 &  --   & 0.55\\
	Jan 14 2003 & $2.13$  & 800 & 21.80 &  --   & 0.57\\
	Jan 13 2003 & $2.17$  & 800 & 21.76 &  --   & 0.51\\
	\hline
      \end{tabular}
  \end{minipage}
\end{table}

\begin{table*}
  \begin{minipage}{\textwidth}
    \caption{Spectroscopy observation log.}
    \label{tab:spec}
    \begin{tabular}{@{}lccccccl@{}}
      \hline
      Date   & Filter & Spectral domain & Slit width &  $R$\footnote{Spectral resolution from ISAAC manual.}  & Integration &    PA\footnote{Position angle on the sky, convention is positive from North to East.} & Notes on quality and spectral features    \\ 
      &      &  (\mum)   & (arcsec)  &  & (sec) & (degrees) &\\
      \hline
      Apr 04 2002 & $SL$ & 2.80 -- 4.00 & 0.6 &  600  & 24$\times$0.94 & -57.10 & Good. No PAHs.\\
      Apr 04 2002 & $SK$ & 2.25 -- 2.37 & 0.3 &  8900 & 2$\times$300 & +107.25 & Good. CO band-heads\\
      Feb 09 2003 & $SK$ & 2.11 -- 2.23 & 0.3 &  8900 & 2$\times$300 & +107.25 & Good. CO band-heads\\
      Feb 09 2003 & $SK$ & 2.11 -- 2.23 & 0.6 &  4400 & 2$\times$300 & -57.65  & Bad, fringing\\
      Apr 14 2003 & $J$  & 1.25 -- 1.30 & 0.6 &  5200 & 2$\times$600 & -57.65  & Regular. Resolved [Fe\,{\sc ii}] emission\\
      May 23 2003 & $J$  & 1.25 -- 1.30 & 0.3 & 10500 & 2$\times$300 & -37.55, +17.25, +76.65 & Good. Resolved [Fe\,{\sc ii}] emission for PA-37.55\\
      May 23 2003 & $SK$ & 2.11 -- 2.23 & 0.3 &  8900 & 2$\times$500 & -37.55, +17.25, +76.65 & Regular. Resolved H$_2$ emission for PA-37.55\\
      \hline
    \end{tabular}
  \end{minipage}
\end{table*}

We observed NGC\,4438 with the VLT ISAAC imager and spectrograph
\citep{isaac}, which has a field of view of $152\arcsec \times
152\arcsec$ and a pixel scale $0.148\arcsec~\mathrm{pixel}^{-1}$. Our
data consist of the following broad and narrow band (NB) filters: $J$,
$H$, $K_\mathrm{s}$, NB$1.26~\mu$m, NB$1.28~\mu$m, NB$2.07~\mu$m and
NB$2.13~\mu$m filters. Our observing strategy involved acquiring 3 to
10~s exposures in a dithered pattern consisting of four frames with
vertical and horizontal offsets of $\sim 22 \arcsec$ and
$\sim9\arcsec$, respectively.

\begin{figure}
  \centering\includegraphics[bb=67 34 536 512,width=\columnwidth]{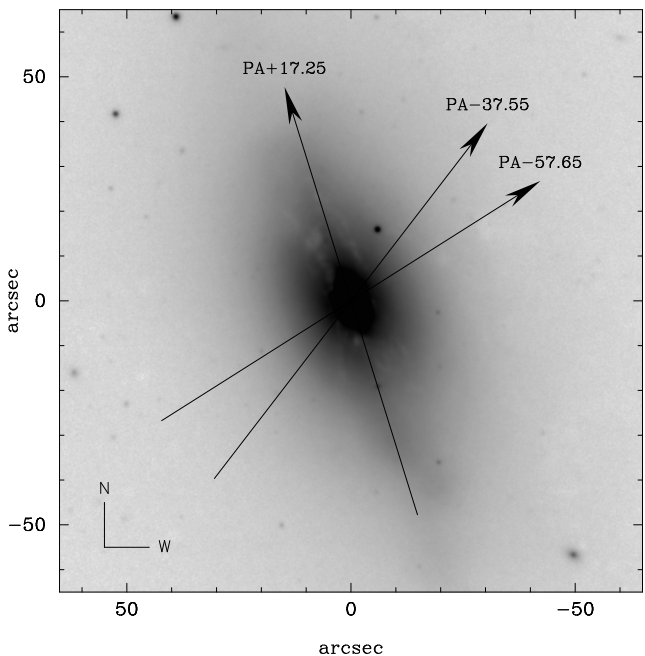}
  \caption{Overlay of the ISAAC 120-arcsec slit on the $J$-band ISAAC
    image of NGC\,4438 for position angles: +17.65, -37.65,
    -57.65. Convention is positive position angle from North to
    East. The $x$- and $y$-axes are east and north offsets in arcsecs
    from the nucleus at J2000 RA $\mathrm{12^h~27^m~45.6^s}$ and J2000
    DEC $\mathrm{+13^{\circ}~00'~32''}$ (from \textit{HST}
    astrometry).}
\label{fig:PAs}
\end{figure}

Data reduction and calibration procedures are summarized in
Appendix~\ref{dr}. Table~\ref{tab:calib} gives the dates, integration
times and calibration parameters for each
observation. Fig.~\ref{fig:PAs} shows the ISAAC $J$-band image of
NGC\,4438 after flat-fielding, bad-pixel correction, and sky and bias
subtraction.

Only the central region of the field (inner 40~arcseconds) has a
symmetric point-spread function (PSF). The outskirts of the array show
extended tails due to optical aberrations. In this central region we
report a seeing of $\sim 0.5\arcsec$ for the NB filters, while for the
broadband filters we report a seeing of $0.6\arcsec - 0.8\arcsec$ (see
Table~\ref{tab:calib}).

\subsubsection{Complementary imaging data}

We also observed NGC\,4438 with the TIMMI2 camera on the ESO~3.6m
telescope on UTC 2002-01-01T08:50, with a lens scale of 0.3~arcsec per
pixel. The target was acquired in the NB~11.9~$\mu$m filter by blind
pointing on the HST coordinates of the nucleus, with a $5 \arcsec$ rms
positional uncertainty in a field of view of $90 \arcsec \times 60
\arcsec$. The object is detected in NB~11.9~$\mu$m, which is free of
emission lines. Yet our aligned chopping and nodding strategy, with a
throw of 30$''$ North-South, fits one of the negative images on the
array. The presence of the chopped image confirms our detection. We
enhanced the signal-to-noise ratio of our image by smoothing with a
Gaussian kernel, with a dispersion of 2~pixels, or $0.6 \arcsec$. Flux
calibration was obtained by comparison with HD81797, with a
11.9~$\mu$m flux density of 100.57~Jy. By fitting an elliptical
Gaussian to the standard star we infer an angular resolution of
$0.76\times 0.71\arcsec$ FHWM. After smoothing, the resolution of the
image in Fig.~\ref{timmi} is about $0.96 \arcsec$. We set the
astrometry by tying the centroid of the 11.9~$\mu$m emission to the
HST coordinates for the nucleus.

We complemented the infrared data with recalibrated WFPC2 \textit{HST}
images, broadbands F450W ($B$), F675 ($R$) and F814W ($I$). We
retrieved these data from the HST archive located at the Canadian
Astrophysics Data Centre (CADC). The CADC pipeline recalibrates the
images with up-to-date calibration files. The details of these
observations are described by \citet{ken02}. Since the pixels covering
the nuclear emission were saturated in $R$ and $I$ images, we used
only the $B$-band image to carry out the photometry of the nucleus.

\subsection{Medium-resolution near-IR spectroscopy}

Near-IR spectra of NGC\,4438 were obtained with ISAAC at VLT on 2003
March-April. We acquired medium resolution spectra at central
wavelengths of 1.274~\mum, 2.170~\mum, and 2.310~\mum, and slit widths
of 0.3 and 0.6~arcsec. The list of observations, filters, position
angles, spectral domains, resolutions and slit widths can be found in
Table~\ref{tab:spec}. Fig.~\ref{fig:PAs} shows some of the slit
positions overlaid on the $J$-band ISAAC image.

\begin{figure}
    \centering\includegraphics[bb=81 156 353 594,angle=270,width=\columnwidth]{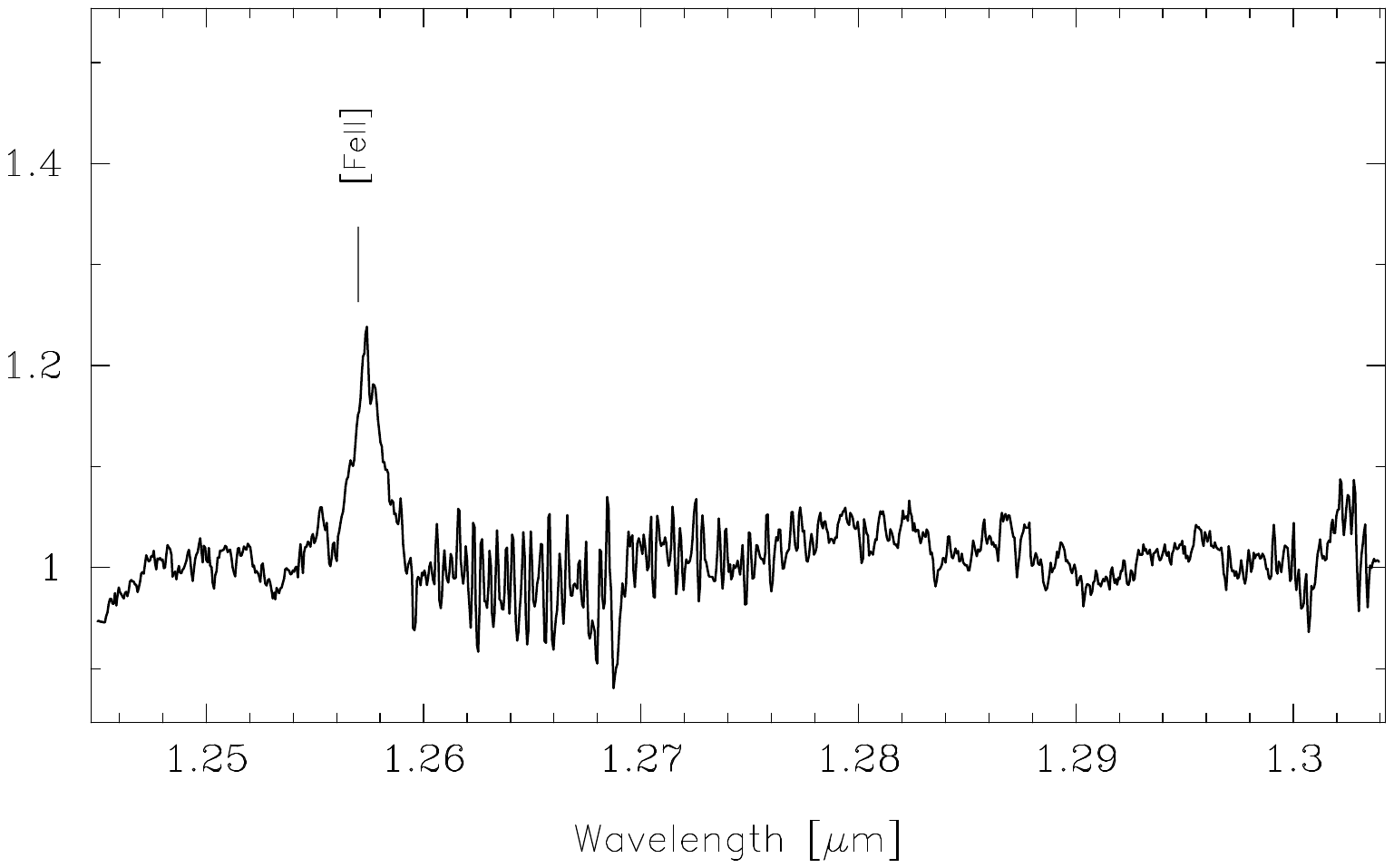}\vspace*{.7cm}
    \centering\includegraphics[bb=81 156 353 594,angle=270,width=\columnwidth]{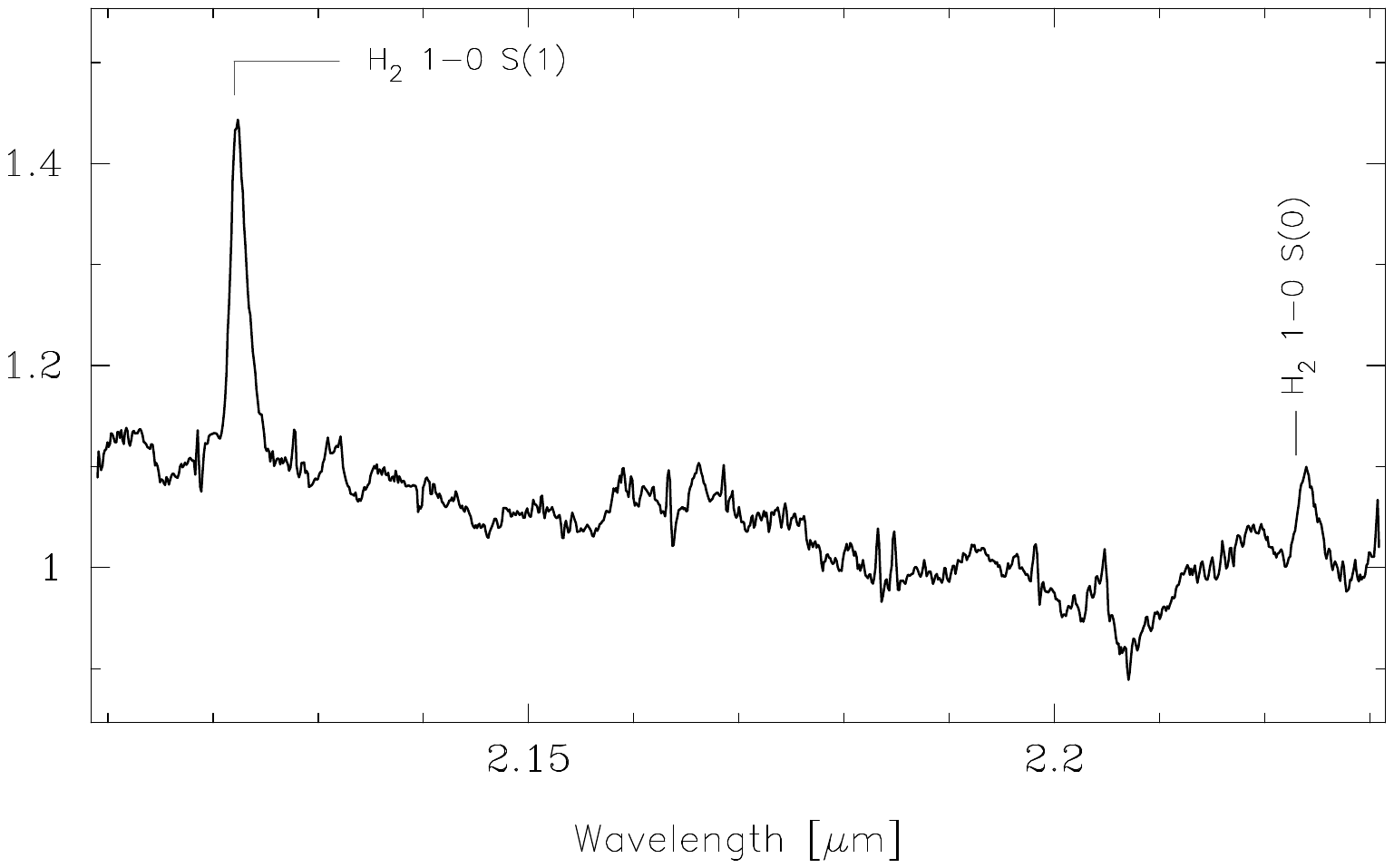}
    \caption{ISAAC near-IR spectra of the nucleus of NGC\,4438. Top
      spectrum is centred on 1.274~\mum, while the bottom spectrum is
      centred on 2.170~\mum. Top and bottom spectra were acquired with
      position angles are -37.55$^{\circ}$ and -57.65$^{\circ}$,
      respectively. Notice the absence of the hydrogen recombination
      lines, Pa$\beta$ at 1.28~\mum\ and Br$\gamma$ at 2.16~\mum, in
      the spectra. Both spectra were extracted from a 0.74~arcsec
      apperture. Wavelengths are in microns, \note{flux densities are
        in units of 6.5$\times 10^{-15}$~W~m$^{-2}$~$\mu$m$^{-1}$.}}
    \label{fig:spec}
\end{figure}

The sky background (including OH skylines) was removed by differencing
along the slit, with nod throws of 50~arcsec (or 30~arcsec in the case
of standard stars, STDs hereafter). Wavelength calibration was
obtained by comparison with arc lamps. The spectra were extracted in
0.74~arcsec centred on the peak of emission. Telluric absorption
spectra and flux density calibration were obtained by observing the
early-type star HD115709 (spectral type A1IV) and comparing with a
black-body spectrum. Stellar absorption features at Pa$\beta$ and
Br$\gamma$ were accounted for by fitting Voigt profiles in a total
wavelength range of 0.015~\mum\ in the vicinity of the stellar
absorption lines. \note{Samples of the reduced spectra, centred at
  1.274~\mum\ and 2.170~\mum, are shown in Fig.~\ref{fig:spec}.}

Flat-fielding was hampered by short-term variations in the CCD pixel
gains. The observatory pipeline flat fields could not suppress the
low-level features on scales of 0.005~\mum\ seen in the spectra of
Fig.~\ref{fig:spec}. The features under $\sim 2\times
10^{-15}$~W~m$^{-2}$~$\mu$m$^{-1}$ were not reproduced in different
exposures, which suggests flat-fielding artifacts. Try as we might we
could not improve on the pipeline flats. We treat the low-level
artifacts as noise, so that the depth of the spectroscopy did not meet
our expectations. The ISAAC near-IR spectroscopy is nonetheless
informative on the kinematics of the CO band-heads, [Fe\,{\sc
    ii}]~1.26~\mum\ and H$_2$~2.12~\mum.


\subsection{Low-resolution spectroscopy}

We acquired a low-resoution L-band (3.8~\mum) spectrum at PA +56.94,
in 7 ABBA cycles, for a total integration time of 1~h (with individual
DITs of 0.935~s and 4 NDITs, a chopthrow of 20~arcsec, and a chopping
frequency of 0.11~Hz). The spectrum shows a continuum rising towards
longer wavelengths. This L-band spectrum does not show any spatially
resolved features. The flux density in the collapsed 0.6~arcsec slit
is $3\times 10^{-15}$~W~m$^{-2}\mum^{-1}$ at 2.8~\mum, and $6\times
10^{-15}$~W~m$^{-2}\mum^{-1}$ at 4.0~\mum. The noise varies as a
function of wavelength, between $10^{-15}$ to
$10^{-14}$~W~m$^{-2}\mum^{-1}$. This spectrum shows no evidence of the
presence of polycyclic aromatic hydrocarbons (PAHs) emission in
NGC\,4438.

\section{Results}\label{results}

\subsection{Extinction estimates}\label{extin}

\begin{figure}
  \centering\includegraphics[bb= 91 92 562 568,width=\columnwidth]{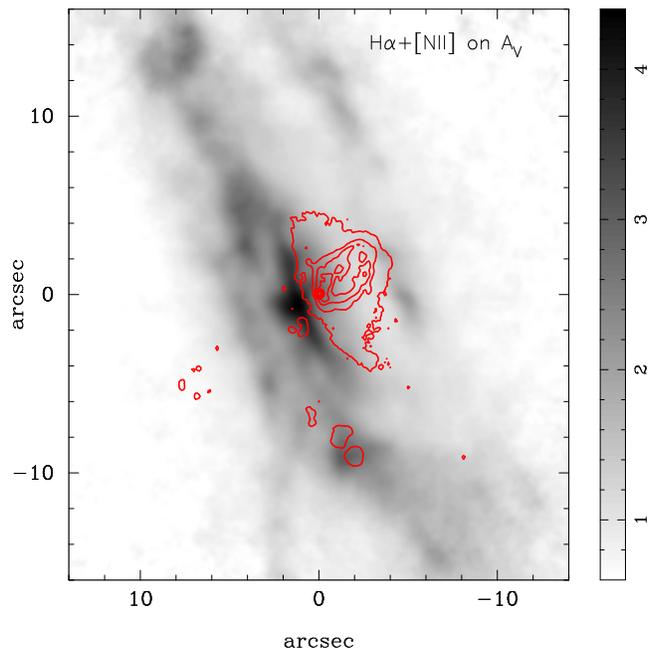}
  \caption{Extinction in the central 40\arcsec\ of NGC\,4438,
    superposed on \textit{HST} H$\alpha$+[{\sc N\,ii}] contour map
    \citep{ken02}. The highest extinction corresponds to the region
    surrounding the nucleus, with a peak of $A_{\rm V}\sim4.4$~mag at
    around 1\arcsec east of the nucleus. Contours are -0.01, 0.06,
    0.27, 0.62, 1.1, 1.73, 2.49, 3.4, 4.44, 5.62, 6.94 times
    $2.4\times 10^{-16}~\mathrm{erg~s^{-1}~cm^{-2}}$. The $x$- and
    $y$-axes are east and north offsets in arcsecs from the nucleus at
    J2000 RA $\mathrm{12^h~27^m~45.6^s}$ and J2000 DEC
    $\mathrm{+13^{\circ}~00'~32''}$ (from the \textit{HST}
    astrometry).}
  \label{AV}
\end{figure}

A consequence of the dependence of interstellar extinction on
wavelength is that the near-IR imaging often reveals surprising
morphological differences when compared to optical observations
\citep[see ][ for a description of the Galactic extinction
  law]{car89}. Nuclear extinction estimates are very uncertain in
NGC\,4438. The Balmer decrement $\mathrm{H\alpha/H\beta}$ gives
$A_{\rm V}=3.4$ \citep[using the extinction law from][]{car89}, but
this represents a luminosity-weighted average over the central $2''
\times 4''$ extraction aperture of the low-resolution ground-based
spectroscopy reported by \citet{ho97}, and also a lower limit to the
true extinction.

In order to further investigate the spatial distribution of the dust
in the system we generated an extinction map (see Fig.~\ref{AV}). This
was done by transforming the $J-K$ colour image into a colour excess
map, $E(J-K)$. We assumed a flat and constant stellar population, as
expected for bulges of spirals, with $E(J-K)=0.8$ (calculated over an
$7\arcsec \times 7\arcsec$ aperture in a region devoid of
structure). The colour excess map $E(J-K)$ and the column of hydrogen
nuclei $N_{\rm H}$ are related by $N_\mathrm{H}/E(J-K) = 1.1 \times
10^{22}~\mathrm{cm^{-2}\, mag^{-1}}$, assuming a Galactic gas-to-dust
ratio and $R_V = 3.1$ \citep{allens}. We inferred a mean colour excess
$E(J-K)=0.40$ for an aperture matched with the $2\arcsec \times
4\arcsec$ used by \citet{ho97}. Using the standard IR interstellar
reddening law \citep{car89}, it implies a visible colour excess of
$E(B-V)=0.7$, which is fully consistent with the colour excess
reported by \citet{ho97}, of $E(B-V)=0.61$. As revealed by the
presence of strong dust lanes, the nuclear extinction is highly
variable toward the central hundred parsecs \citep{ken02}. 

The spatial distribution of extinction and dust is shown in
Fig.~\ref{AV}. The lower edge of the H$\alpha$ bubble emission is
coincident with bands of strong extinction. These region of high
extinction seem to be surrounding the nucleus (given by the peak of
H$\alpha$ emission), with a peak of $A_{\rm V}\sim 4.4$~mag at around
1\arcsec east of the nucleus.

The measured colour excess implies an average hydrogen column density
$N_\mathrm{H}= 2.7 \times 10^{21}~\mathrm{cm^{-2}}$ at the base of the
NW bubble (shown in Fig.~\ref{AV}), and a mean column $N_\mathrm{H}=
1.2 \times 10^{21}~\mathrm{cm^{-2}}$ for the rest of the NW bubble
\citep[using the total to selective absorption from][]{allens}. This
column density is in reasonable concordance with the values reported
by \citet{mac04}. \citet{mac04} obtained a best-fitted column density
of $\sim 2 \times 10^{21}~\mathrm{cm^{-2}}$ from modelling of the
X-ray spectra of the NW bubble. The uncertainties in our extinction
estimates are mainly due to variations in the stellar population near
the nucleus.

\begin{figure*}
  \includegraphics[bb=64 29 562 580,angle=270,width=.9\columnwidth]{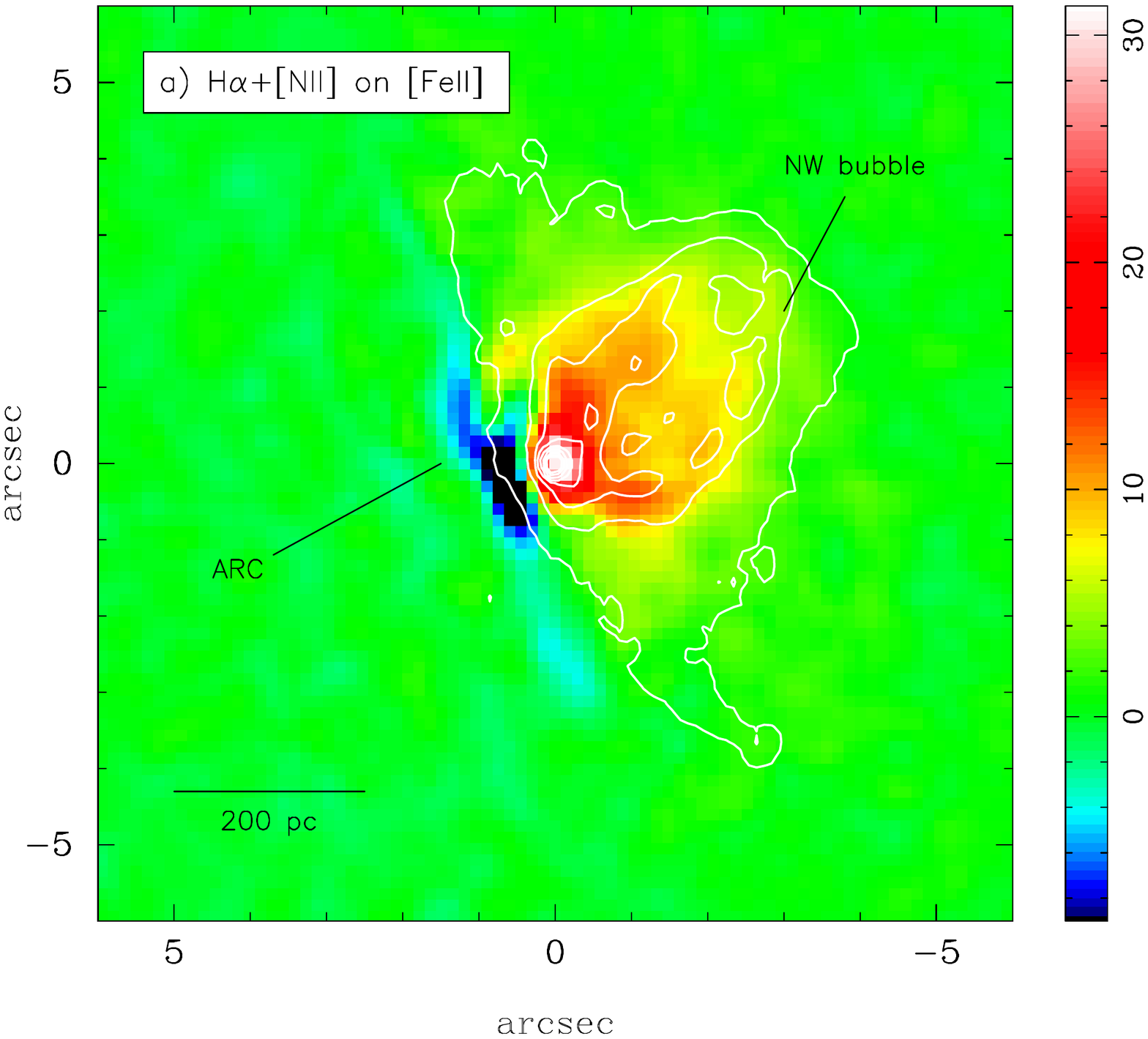}\hspace*{1cm}
  \includegraphics[bb=64 164 562 715,angle=270,width=.9\columnwidth]{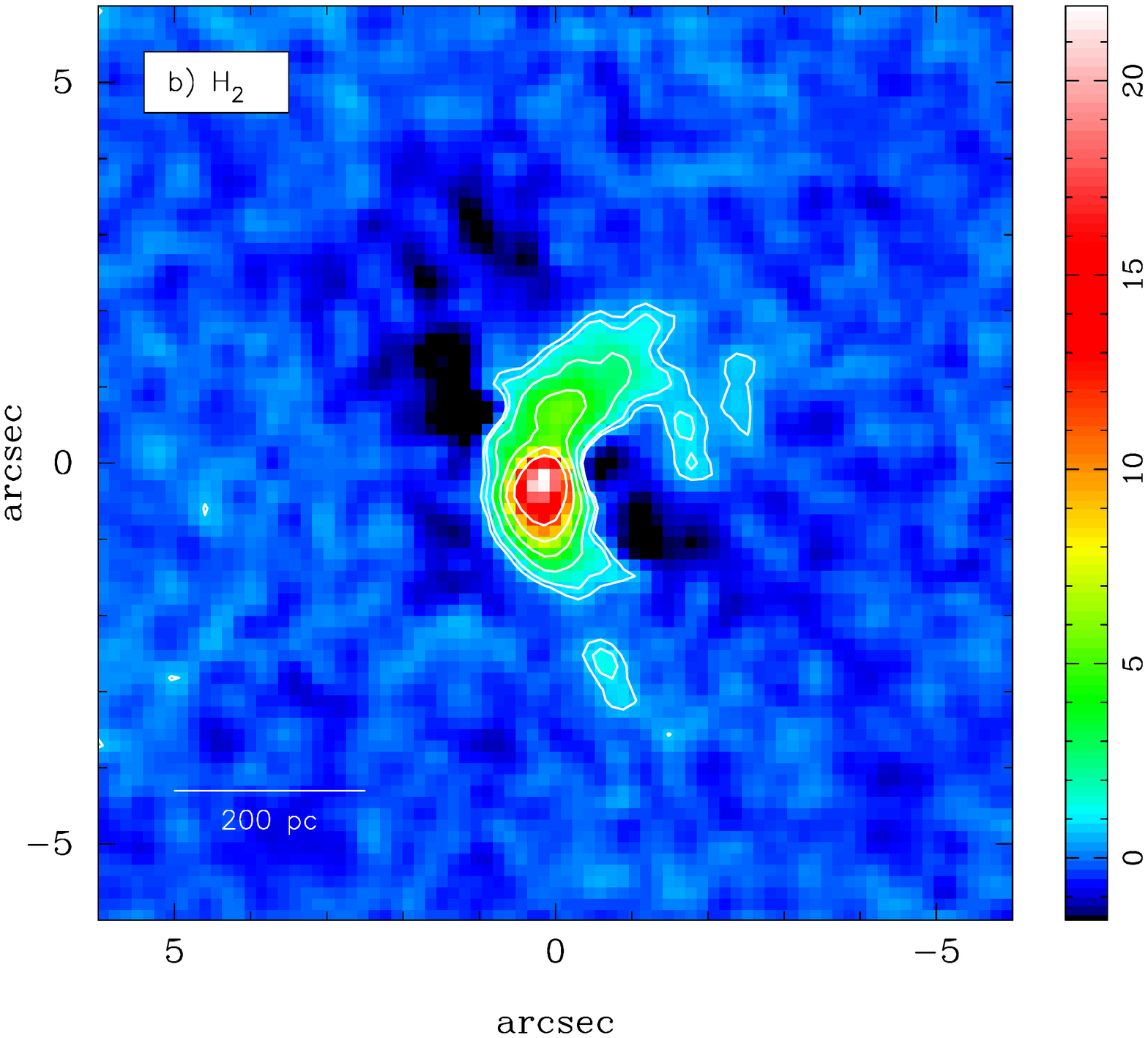}
  \caption{a) ISAAC [Fe\,{\sc ii}] continuum subtracted emission-line
    map of the central 10\,\arcsec\ in NGC\,4438. Note that there is a
    negative arc due to Pa$\beta$ contamination in our continuum
    frame. The colour-scale is exponential and the contours correspond
    to H$\alpha$+[{\sc N\,ii}] emission at 0.12, 0.14, 0.18, 0.27,
    0.39 and 0.55 times 32.9~MJy~sr$^{-1}$. b) Molecular hydrogen
    (${\rm H}_2$) in the central 10\arcsec\ of NGC\,4438. The
    colour-scale is exponential and the contours are 0.04, 0.09, 0.18,
    0.31, 0.45 and 0.62 times 21.9~MJy~sr$^{-1}$. The $x$- and
    $y$-axes are east and north offsets in arcsecs from the nucleus.}
    \label{fig:bub}
\end{figure*}

\begin{figure*}
  \includegraphics[bb=64 29 566 580,angle=270,width=.9\columnwidth]{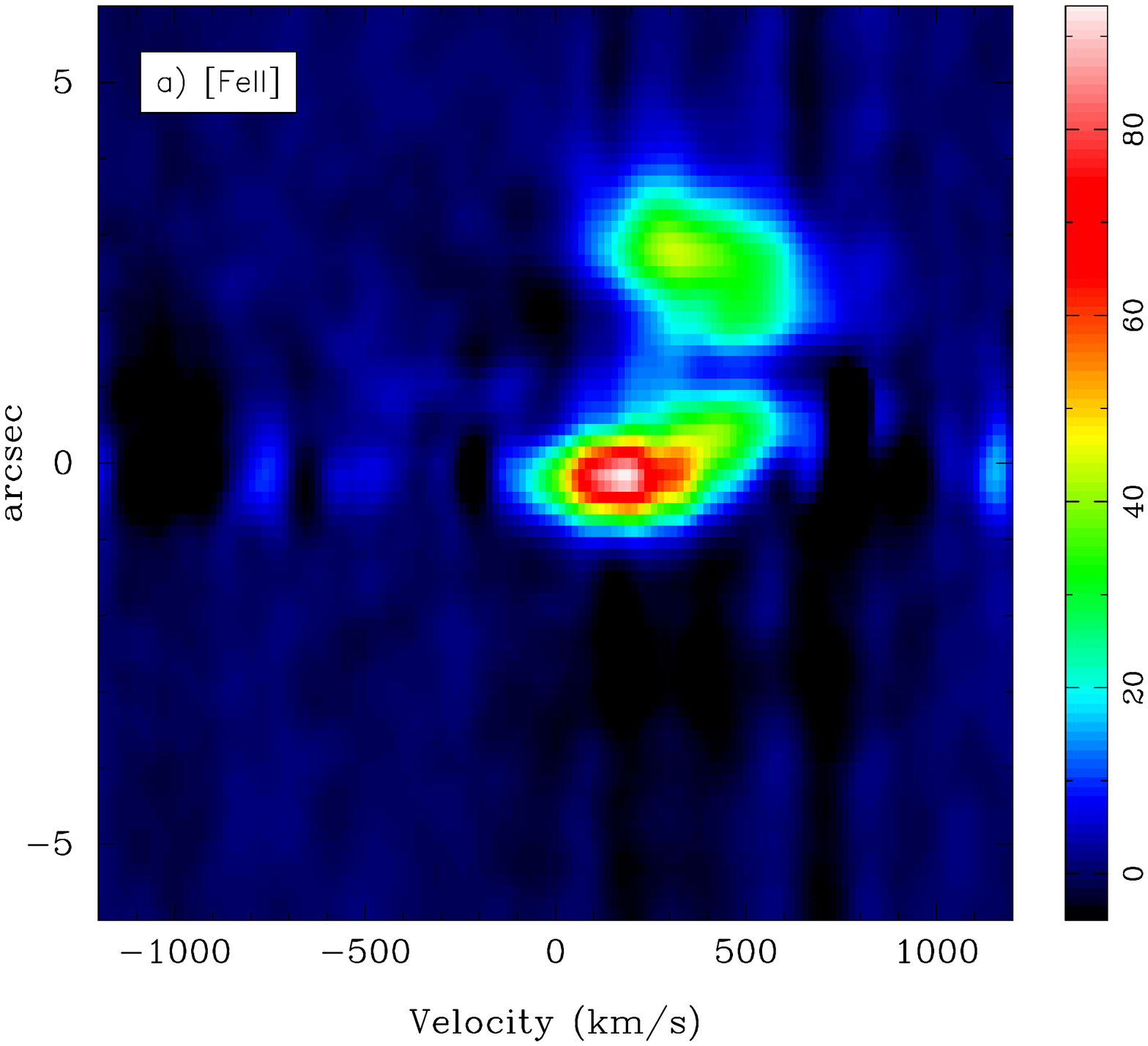} 
  \hspace*{1cm}
  \includegraphics[bb=64 29 566 580,angle=270, width=.9\columnwidth]{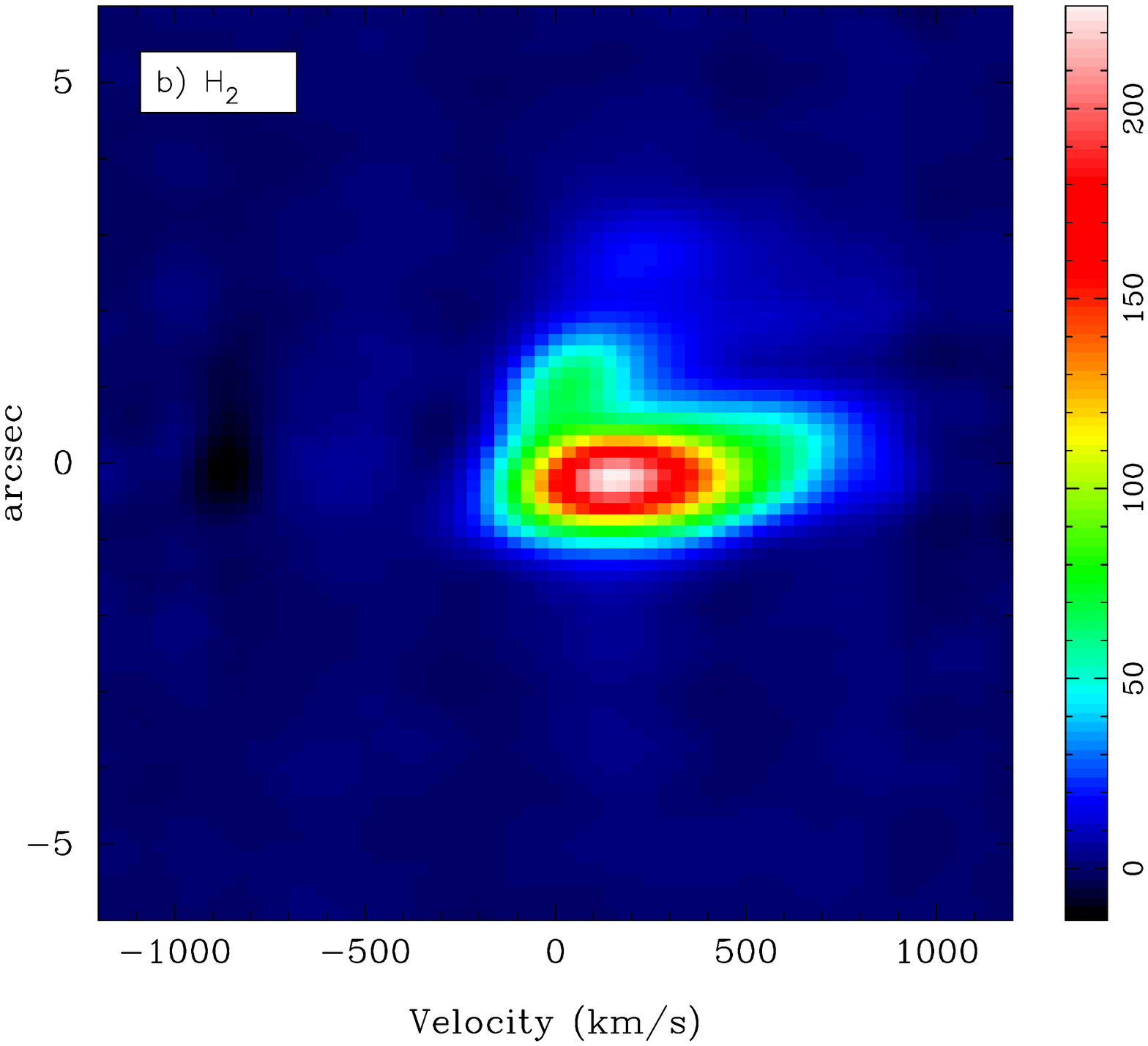}
  \caption{Results of long-slit observations of NGC\,4438. Panel a)
    shows [Fe\,{\sc ii}] at 1.257~$\mu$m extended emission while panel
    b) corresponds to ${\rm H}_2$ at 2.122~$\mu$m. Spectral PSF
    subtraction was carried out in order to remove the background and
    continuum of the galaxy. The slit position for both spectra
    corrresponds to the direction along the major-axis of the NW
    bubble (position angle is -57.65 degrees), thus giving the
    velocity gradient across the shell. Intensity is given in units of
    the rms noise on a pseudo-colour exponential scale. The $y$-axis
    is the spatial offset along the slit in arcsec, and the $x$-axis
    shows the velocity in the rest frame of the object in \kms.}
  \label{fig:extended}
\end{figure*}

\subsection{Maps of extended emission}\label{sec:nuc}

Fig.~\ref{fig:bub} shows the [Fe\,{\sc ii}] $1.26~\mu$m and H$_2$
2.122~$\mu$m maps of line emission, after subtracting the continuum
emission. The fact that there is some negative emission at the base of
the NW bubble implies that there is some Pa$\beta$ emission
contaminating the continuum image. We are certain that this
contamination does not affect the NW extended emission since the
spectrum shows no trace of Pa$\beta$ emission \note{at a noise level
  of 1\,$\sigma=1.06 \times 10^{-16}$~W~m$^{-2}$~$\mu$m$^{-1}$,} when
the slit is aligned with the NW bubble (PA -37.55 and PA -57.65, see
Fig.~\ref{fig:spec}). In the case of H$_2$ emission we used a scaled
version of the $K_{\rm s}$ broadband image to subtract the
continuum. The presence of this strong [Fe\,{\sc ii}] emission is
consistent with LINER spectroscopic surveys \citep{lar98}, confirming
the LINER classification of the NGC\,4438's nuclear region.

The outflow shells are very asymmetric in the optical lines and the
radio and X-ray continuum.  This provides further evidence that the
outflow shells are intrinsically different, and that the observed
dissimilarities are not due principally to extinction. In the near-IR
we have detected only the NW outflow, with no trace of the
south-eastern outflow neither in [Fe\,{\sc ii}] nor H$_2$ emission.

\subsubsection{[Fe\,{\sc ii}] extended emission}\label{sec:feii}

The emission-line map shown in Fig.~\ref{fig:bub}a indicates that
[Fe\,{\sc ii}] emission is coincident with \textit{HST}
H$\alpha$+[{\sc N\,ii}] emission-line detected by \citet{ken02},
\textit{Chandra} X-ray emission reported by \citet{mac04}, and the
complex radio continuum blob A described by \citet{hum91}. Emission
from a counter-shell in the southeastern (SE) region is clearly
detected in H$\alpha$+[{\sc N\,ii}] and radio \citep[see ][]{ken02},
but is much fainter, and is located much further from the nucleus
(9\arcsec vs 4\arcsec) than the much-brighter NW
shell. Fig.~\ref{fig:bub} also reveals a lack of this SE
emission. This is because the SE emission is 15 times fainter than the
NW component in H$\alpha$+[{\sc N\,ii}] \citep{ken02}, and even weaker
in the X-ray \citep[a factor of 32 lower counts than the NW
  shell,][]{mac04}. Also, most of the bubble-like emission seen in
H$\alpha$+[{\sc N\,ii}] corresponds to [{\sc N\,ii}] emission since
H$\alpha$ is about two times weaker than ionized nitrogen emission
according to Fig.~\ref{fig:ha_nii}. The bright {\sc H\,ii} regions,
detected in H$\alpha$+[{\sc N\,ii}] emission 8\arcsec\ toward the
south of the nucleus by \citet{ken02}, do not appear neither in
Pa$\beta$ or [Fe\,{\sc ii}] emission nor in X-ray emission
\citep{mac04}.


In AGN galaxies, the [Fe\,{\sc ii}] emission is thought to arise from
the region radiating narrow line emission \citep{mou00, rif06}. Such
regions in AGN can be produced either as a result of photoionization
by a nuclear source, including X-ray heating from the central AGN, or
via shock excitation by radio jets \citep{rod05, rif06}. The dominant
excitation mechanism of the [Fe\,{\sc ii}] emission is still under
debate. \citet{sim96} have argued that photoionization is the dominant
excitation mechanism of [Fe\,{\sc ii}] and that shocks by radio jets
account for only about 20 per cent of the emission in Seyfert
galaxies. \citet{rod05} have shown that the [Fe\,{\sc ii}]/Pa$\beta$
ratio is a good indicator of the relative contribution of
photoionization and shocks, since [Fe\,{\sc ii}] seems to be more
tightly correlated with the radio emission than hydrogen recombination
lines in AGN \citep{sim96}.

Radio maps of NGC\,4438 presented by \citet{hum91} show extended
emission, with a shell-like morphology, that is well aligned with
H$\alpha$ and [{\sc N\,ii}] \citep[see figures~3 and~4 in][]{ken02},
and therefore also spatially coincident with [Fe\,{\sc ii}]. This
morphological correlation, along with the non detection of extended
Pa$\beta$ emission in our spectra of the NW bubble, may imply that
shock excitation by radio jets contributes to an important fraction of
the [Fe\,{\sc ii}] emission in NGC\,4438.

\subsubsection{H$_2$ extended emission}\label{sec:h2}

Molecular hydrogen emission is present in many host galaxies of AGNs
\citep{rod05,rif08,rif09}. In the traditional AGN picture, a dusty
torus shields the H$_2$ gas from the dissociating AGN radiation
field. The H$_2$ emission in NGC\,4438 seen in Fig.~\ref{fig:bub}b
could also be associated with the central AGN engine but it does not
seem to trace the NLR gas \citep{rod05}. As it can be seen in
Fig.~\ref{fig:bub} the molecular hydrogen is distributed only in the
inner few hundred parsecs. The H$_2$ emission is much stronger on the
north side of the bubble than on the south side. Fig.~\ref{fig:bub}b
shows that most of the molecular hydrogen gas is concentrated toward
the circumnuclear region tracing the presence of a possible dusty
torus, as in the case of NGC\,3727 \citep{rod05} or NGC\,4051
\citep{rif08}. Also, some H$_2$ emission extending towards the NW
bubble can be seen in our observations.

The position-velocity maps depicted in Fig.~\ref{fig:extended} show
that the kinematics of the [Fe\,{\sc ii}] and the H$_2$ emitting gas
are distinct. Furthermore, the nuclear [Fe\,{\sc ii}] line is broader
than the H$_2$ emission. As measured from the nuclear spectra shown in
Fig.~\ref{fig:spec}, the full-width at half-maximum (FWHM) of the
[Fe\,{\sc ii}] and H$_2$ $\lambda$2.12\mum\ lines are 415 and
270~\kms, respectively. This implies that the [Fe\,{\sc ii}] is
originating from a kinematically more disturbed gas than the H$_2$
emitting gas, which is in agreement with previous observations of
other active galaxies, such as NGC\,2110 \citep{sto99} and NGC\,4051
\citep{rif08}. A possible interpretation for this is that the H$_2$
emitting gas is more restricted to the galactic plane, perpendicular
to the radio jets, while the [Fe\,{\sc ii}] emitting gas extends to
higher latitudes from the galactic plane \citep{rif08,sto99}. These
results suggest that the H$_2$ excitation is likely to be dominated by
X-ray heating from the central AGN.

\subsubsection{Pa$\beta$ extended emission}\label{sec:pab}

Fig.~\ref{fig:bub}a also shows a negative arc to the east of the
nucleus, perpendicular to the NW shell. In this arc Pa$\beta$ is
stronger than the [Fe\,{\sc ii}] emission. The distance between the
nucleus and the arc is about 0.5\arcsec\ ($\sim$35~pc), which is much
larger than the values pointed out for the radius of the putative
torus. These values are typically smaller than 5~pc \citep[see, for
  example,][]{jaf04,min04,rif09}. A possible interpretation for this
emission is that it might be associated with circumnuclear star
formation in the outer parts of the torus.

The circumnuclear region seen in negative in Fig.~\ref{fig:bub}a has a
Pa$\beta$ flux density of 0.5~mJy (integrated over the arc-shaped
feature), which corresponds to a Pa$\beta$ luminosity of $3.6 \times
10^{40}~\mathrm{erg~s^{-1}}$. If this emission has a starburst origin
and we assume solar abundances, and a Salpeter initial mass function
$\psi \propto M^{-2.35}$ with an upper mass cutoff of 100~\Msun, we
can estimate the star formation rate (SFR) in the arc
feature. \citet{ken98} find by extrapolation: $
\mathrm{SFR}(\Msun~\mathrm{yr^{-1}}) = 7.9 \times 10^{-42}
L(\mathrm{H}\alpha)~(\mathrm{erg~s^{-1}})$, computed for Case B
recombination at $T_e = 10^4$~K. Under these conditions the SFR as a
function of Pa$\beta$ luminosity is,

\begin{equation}
  \mathrm{SFR}\left (\Msun~\mathrm{yr^{-1}}\right) = 1.4 \times
  10^{-41}~
  L\left(\mathrm{Pa}\beta\right)~\left(\mathrm{erg~s^{-1}}\right),
\end{equation}

for $j_{\mathrm{H}\alpha}/j_{\mathrm{Pa}\beta} = 17.6$, where $j$ is
the emissivity of the line \citep{sto95}. The extinction-corrected
nuclear star formation rate, using a central extinction $A_{\rm V}$ of
4~mag, is 1~\Msun~yr$^{-1}$. It represents an overestimated SFR in the
circumnuclear region because this emission could include a
contribution from the AGN, due to gas that has been ionised by an
accreting black hole. For a global description of the star formation
history of NGC\,4438 as a whole see \citet{bos05}.

\subsection{CO-band stellar kinematics}\label{sec:CO}

\begin{figure}
  \centering\includegraphics[bb=25 340 557 718,width=\columnwidth]{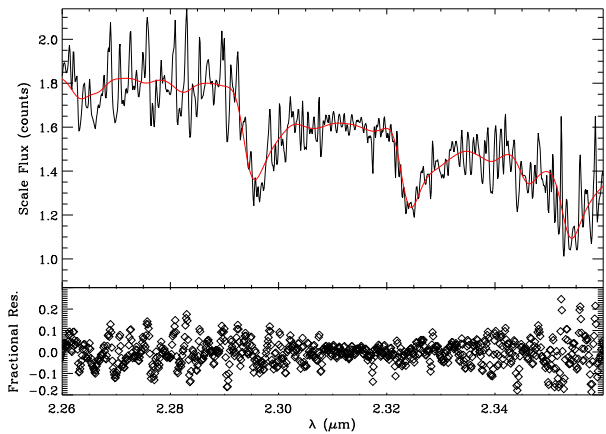}
  \caption{Medium-resolution spectrum centred at 2.310~$\mu$m with a
    slits of 0.74 arcsec. The red line represents the best galaxy
    model fitted by pPXF (see text) as a linear combination of
    template spectra.}
    \label{fig:co_spec}
\end{figure}

\begin{figure}
    \centering\includegraphics[bb=53 359 557 718,width=\columnwidth]{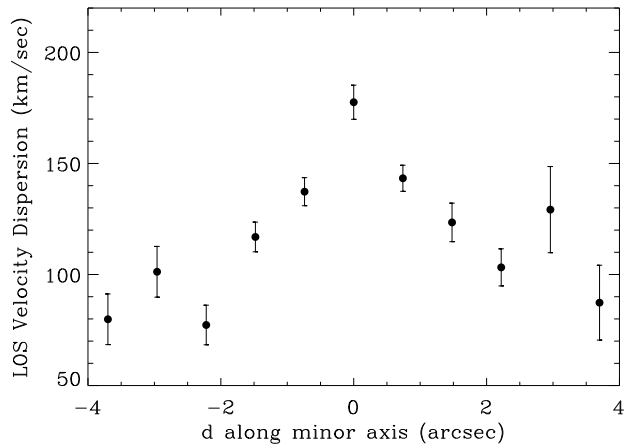}
    \caption{Line-of-sight (LOS) stellar velocity dispersion
    ($\sigma$) as a function of distance along the minor-axis of the
    galaxy. The LOS velocity dispersion reaches a maximum at the
    nucleus of the galaxy with $\sigma_{\rm max} \sim 180~\kms$}.
    \label{fig:vel}
\end{figure}

NGC\,4438 is an inclined large spiral galaxy and its kinematics are
undoubtedly difficult to study mainly due to the multiplicity of
components that build up spiral galaxies. Moreover, NGC\,4438's highly
disturbed morphology and the presence of dust make the inference of
kinematic measurements a difficult task. On the positive side, its
inclined disc permits a relatively simple identification of the
direction which might be expected to define one of the principal axes
of the bulge.

CO-band stellar line-of-sight kinematics along the minor axis (PA =
+107.5$^{\circ}$) were derived by fitting the galaxy spectra (see
Fig.~\ref{fig:co_spec}) with a linear combination of template spectra
chosen from the Gemini Near-IR spectral templates library\footnote{see
  near-IR resources at:\newline
  \texttt{http://www.gemini.edu/sciops/instruments/}}. Stellar
templates were rebinned to the spectral resolution of the galaxy
spectra (15.47 \kms\ pixel$^{-1}$). The best-fitting parameters were
determined by $\chi^2$ minimisation in the pixel space using the
penalized pixel fitting method (pPXF) developed by \citet{cap04},
which required a signal-to-noise ratio of at least 30 in order to get
reliable stellar kinematics. Errors were estimated by using a
Monte-Carlo scheme. They were obtained as the standard deviation of
the kinematical parameters ($V_{\rm los}$ and $\sigma_{\rm los}$) for
many realisations ($N =$~300) of the input spectra by adding Gaussian
noise to the best model of the galaxy spectrum. The kinematical study
of the nature of this putative central mass object requires detailed
kinematical model \citep[e.g., ][]{cre99,cap09}. Unfortunately, the
scarcity of our data does not allow us to perform this kind of
analysis.


The results shown in Fig.~\ref{fig:vel} that NGC\,4438 exhibits a
strong peak of the line-of-sight velocity dispersion along the minor
axis on the centre of the galaxy. The velocity dispersion varies from
100 to 180~\kms\ in a region of about 3\arcsec ($\sim$ 230~pc) in
size, which could suggest the presence of a massive central object.

The radius of the sphere of influence of this possible central massive
object is just $\sim$0.1\arcsec, which is considerably smaller than
our 0.74\arcsec\ spatial resolution. Therefore, the enhancement
observed in the stellar velocity dispersion curve in the inner
$\sim$230~pc, where sigma rises from about 100 up to 177~\kms, might
be only due to the galaxy bulge potential. However, the sphere of
influence is not a hard limit to the influence of a black hole on its
host bulge. The influence of the compact object will still be felt at
a larger radii but to a lesser extent. Comparing Fig. 8 with the
dispersion profiles of other AGNs such as NGC\,2549, whose sphere of
influence is also $\sim$0.1\arcsec\ \citep[see fig. 4 in][]{kra09},
NGC\,4438's velocity dispersion seems to be even steeper in the
central couple of arcsec. Therefore, we cannot rule out the presence
of a supermassive black hole as responsible for the central
enhancement in the velocity dispersion profile of NGC\,4438.


\section{\note{Surface brightness modelling of NGC\,4438}}\label{pho}

\subsection{Model description: general considerations}

Does the nucleus seen in our broadband images correspond to a cuspy
surface-brightness profile, or is it a point source?  Our approach to
addressing this question consists in modelling the stellar component
by fitting the bulge with a $r^{1/n}$ S\'ersic law, and the outer
region (disc) with an exponential profile. The nuclear component will
be {\it a priori} represented by a point source at the centre of the
galaxy.  The main advantage of this approach is that it lets us
identify a nucleus over the stellar contribution, whereas a
non-decomposition approach would try to reproduce all the profile
without giving back information about separate components. The
universality of the S\'ersic plus exponential disc decomposition
allows us to compare with other objects. A detailed description of the
surface brightness model, as well as the optimisation algorithm can be
found in Appendix~\ref{sbm}.

\begin{table*}
  \begin{minipage}{\textwidth}
    \begin{center}
      \caption{Best NGC\,4438 galaxy structural parameters.}
      \label{bestfit}
      \begin{tabular}{@{\extracolsep{\fill}}lcccccccccccccccc}
	\hline 
	Filter&$\Delta $&$q$&$c$&$\alpha$&$n$&$R_e$&$I_e$&$R_h$&$I_0$& $n'$& $R_e'$& $I_e'$ & $L_\mathrm{tot}$ & $\log{\mathcal{M}_\mathrm{bh}}$& $\log{\mathcal{M}_\mathrm{bh}(n)}$ \\
	(1)     &(2)    &(3)    &(4)    &(5)    &(6)    &(7)    &(8)    &(9)    &(10)   &(11)   &(12) & (13) & (14)& (15)& (16)\\
	\hline
	\multicolumn{16}{c}{bulge + disc + S\'ersic nucleus}\\
	$J$	&30	&0.57	 &1.10	 &68.3	 &1.65	 &3.88	 &71.76	 &13.46	 &42.65	 &1.75	 &$\sim0$	&572.63	&27.7	&7.19 &6.90$\pm 0.45$\\
	$H$	&30	&0.57	 &1.14	 &68.4	 &1.47	 &3.53	 &94.76	 &12.77	 &57.08	 &1.93	 &$\sim0$	&502.12	&21.8	&6.97 &6.68$\pm 0.45$\\
	$Ks$	&30	&0.57	 &1.13	 &68.7	 &1.52	 &3.49	 &73.39	 &12.86	 &41.34	 &1.71	 &$\sim0$	&570.92	&13.0	&6.66 &6.75$\pm 0.45$\\
	$B$	&30     &0.50	 &1.20	 &68.9	 &0.80	 &2.50	 &23.00	 &8.70	 &30.00	 &1.70	 &0.20	        &154.00	&7.56	& --   & --\\
	\hline
      \end{tabular}
      \end{center}
      Notes.-- (1) Filter name. (2) Angular size of the fitted region.
      (3) \& (4) generalised ellipticity parameters. (5) Inclination
      angle in degrees. (6), (7), (8), (9) and (10) are the S\'ersic
      structural parameters for the bulge, $R_e$ and $R_h$ are in
      units of arcseconds. (11), (12) and (13) are the S\'ersic
      structural parameters for the nuclear source. (14) Total bulge
      luminosity in units of 10$^{42}$~erg~s~$^{-1}$, calculated from
      Equation~\ref{Ftot}. (15) Black hole masses obtained from the
      $\mathcal{M}_\mathrm{bh}$-luminosity density relation
      \citep{mar03}. (16) Black hole masses obtained from the
      $\mathcal{M}_\mathrm{bh}$-$n$ relation \citep{gra07}, the errors
      are estimated using the intrinsic uncertainties in the
      \citet{gra07} relation. All radii are given in arcsecs, the
      intensity units are MJy~sr$^{-1}$ and the black hole masses are
      expressed in \Msun.
  \end{minipage}
\end{table*}

\begin{figure*}
  \begin{minipage}{\textwidth}
    \centering\includegraphics[bb=223 62 583 784,angle=270,width=.77\textwidth]{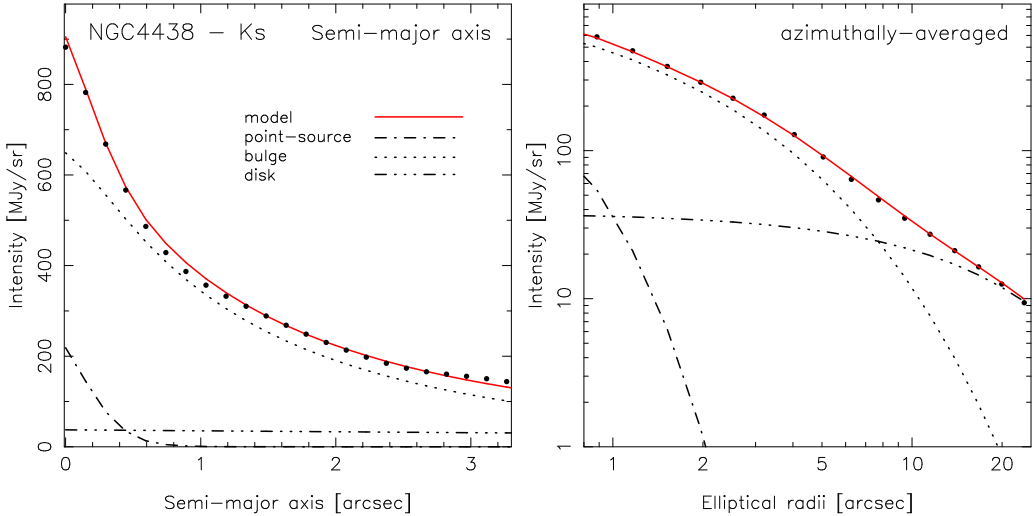}
    \caption{Results of the 2D decomposition of NGC\,4438 (see text).
      \textit{Left:} Semi-major axis cut to the central 3~arcsec. The
      dashed line is the de-reddened galaxy profile. \textit{Right:}
      Surface brightness azimuthally-averaged over a radial elliptical
      annulus (points) of width $\Delta r \sim$0.2~arcsec, with $r$
      defined by Equation~\ref{rad} (axes are in logarithm scale).}
    \label{2dfit}
  \end{minipage}
\end{figure*}

\subsection{Application to the near-IR ISAAC data}

We fitted the nuclear region ($30\arcsec \times 30\arcsec$) of
NGC\,4438 with parametric functions, as described in
Appendix~\ref{sbm}. The model is the sum of three components: a bulge,
a disc and a compact source. The dust features present near the
central region have been taken into account by correcting the images
using the extinction map (see Fig.~\ref{AV}).

The best-fitting structural parameters for each filter are summarised
in Table~\ref{bestfit}. The best fit is always obtained with $n$ close
to 1.7 and $R_e\sim 3$\arcsec\ for the bulge component present at
near-IR wavelengths (see nomenclature in Appendix~\ref{sbm}). The
ellipticity and shape of the isophotes are quite constant with
increasing radius, with an ellipticity $e=0.5$ and slightly disky
isophotes ($c=1.2$, see Appendix~\ref{sbm}).

Since the integrated flux density presented in \citet{pen02} is
undefined in the case of `disky' isophotes, we calculated a general
expression for the integrated S\'ersic profile flux density, in
generalised elliptical geometry, including the case of `disky'
isophotes (Equation~\ref{Ftot}). The integrated luminosities of the
bulge were computed from the flux densities integrated over all radii.

The result of the two-dimensional fit for the $J-$filter is presented
in Fig.~\ref{2dfit}. The left panel shows a cut along the major axis
of the galaxy (dashed curve), the model (solid curve) and its
components: bulge (dotted curve), disc (dash-dot-dot curve) and the
nuclear component (dash-dot curve). That figure shows an excellent
agreement between the model and the galaxy's light distribution. In
this central region we can see a conspicuous point-source standing out
of the bulge. However, galaxies with power-law profiles may show
substantial differences along the major and minor axis, due to
isophote twists or ellipticity. For this reason, the right panel shows
the surface brightness azimuthally-averaged along a radial elliptical
annulus\footnote{With the same nomenclature of the left panel, in
  logarithmic scale}; this panel also shows an excellent match between
model and data at large radii.

We find evidence for a nuclear point-source unresolved at
0.8\arcsec\ resolution (see Fig.~\ref{2dfit}) with $M_K = -18.7$ and
$J-K = 0.69$. Its extinction-corrected near-IR integrated fluxes in a
3\arcsec\ aperture for each band are listed in
Table~\ref{nucphot}. The computed bulge luminosity in each broadband
are listed in Table~\ref{bestfit}. A very similar system was studied
by \citet{pen02}, who found a nuclear point source embedded in
NGC\,4278, which is an elliptical Seyfert galaxy, also with a LINER
nucleus.

\subsection{Application to the \textit{HST}-WFPC2 data}

Since the \textit{HST} $B$ band image shows much more structure than
the near-IR data (due to extinction being more severe at shorter
wavelengths), we followed two different approaches in order to perform
the photometry of the nucleus.\footnote{We did not model the $R$ and
  $I$ band images because the few pixels covering the nucleus were
  severely saturated.}  The first approach was similar to the one used
for the near-IR data, i.e., fitting a S\'ersic bulge plus an
exponential disc and a nuclear source. The nuclear source was
represented by an extra S\'ersic profile, since the central source in
the $B$-band image is clearly resolved. The result was a less
prominent bulge in the optical than in the near-IR. In the optical
case the surface brightness is well represented by a bulge with $n =
0.8$ and a compact source well-fitted by a steeper S\'ersic component
with $n = 1.7$ and $R_e=0.2$\arcsec, i.e,. a resolved nuclear source
of approximately $16$~pc in size. An $n$ value of 0.8 seems very low
for a bulge of a spiral and it may not represent the real bulge
surface distribution, this is probably because of the severe obscuring
effect of dust at those wavelengths.

The second approach consisted of modelling the structure of the
central $2\arcsec \times 2\arcsec$ in the $B$-band image with a base
of legendre polynomials $P_l (x)$ and $P_k (y)$, both at order $l,k =
6$, in order to get accurate photometry of the nucleus. We added a
resolved compact-source, represented by
$F_0\,\delta(x_\mathrm{cen},y_\mathrm{cen})$ convolved with a gaussian
of $\mathrm{FWHM} = 0.4\arcsec$. The free parameters were the
$(l+1)\times(k+1)$ legendre coefficients, the centroid of the
compact-source $(x_\mathrm{cen},y_\mathrm{cen})$, the FWHM of the
nuclear source, and the central intensity $F_0$. The result was
similar to the near-IR case: a resolved compact source with
$\mathrm{FWHM}= 0.4$\arcsec, equivalent to $31~\mathrm{pc}$, appears
to stand out of the nuclear region (see Table~\ref{nucphot}). This
approach is fully consistent with the first approach explained above,
which also yields a compact source with $R_e = 0.2$\arcsec\ since
$\mathrm{FWHM}\sim 2 R_e$.

The difference between the near-IR and optical parameter values can be
explained mainly by the fact that the extinction is more severe in the
optical than in the near-IR. Also, in spiral galaxies the bulge is
mainly composed of late-type stars, hence yielding a more prominent
bulge in the near-IR than in the optical. Photometry of the extracted
nuclear source yields $M_B \sim -12.24$~mag.

\begin{table}
  \begin{minipage}{\columnwidth}
    \begin{center}
      \caption{Near-IR nuclear source photometry.}
      \label{nucphot}
      \begin{tabular}{@{\extracolsep{\fill}}lcccc}
        \hline 
        Filter name & Central $\lambda$  & $F_{\lambda}$  & $M_\lambda$ &  $L_{\lambda}$\\
        & ($\mu$m)  & (mJy)  & (mag)       & ($10^{41}$~erg~s$^{-1}$)\\
        \hline
        $J$        & 1.25       & $10$           & $-17.96$    & $7.35$\\
        $H$        & 1.65       & $12$           & $-18.66$    & $6.67$\\
        $K_{\rm s}$ & 2.16       & $8$            & $-18.67$    & $3.40$\\
        \hline
      \end{tabular}
      \end{center}
      $^a$The specific luminosity was computed as $L_{\nu} = 4\pi D^2
      \nu F_{\nu}$, at a distance $D~=~16$~Mpc.
  \end{minipage}
\end{table}

\section{Discussion}\label{dis}

\begin{figure}
  \centering\includegraphics[bb=81 85 518 518,width=.8\columnwidth]{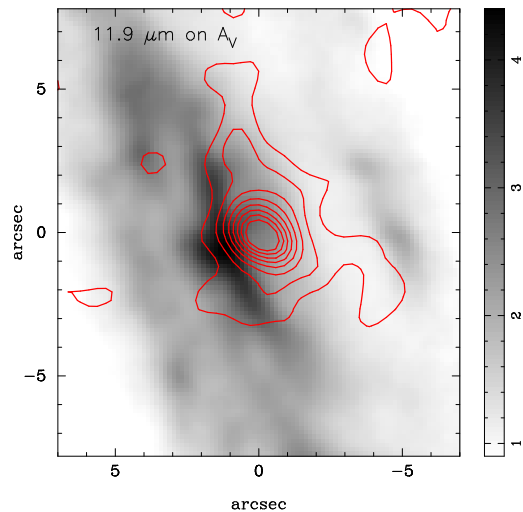}
  \caption{ISAAC $A_{\rm V}$ grayscale map superposed on TIMM2 NB
    11.9~$\mu$m emission map (contours). Contours are -0.27, 0.24,
    0.40, 0.52, 0.61, 0.69, 0.77, 0.83 and 0.89 times
    385~MJy~sr$^{-1}$. The $x$- and $y$-axes are east and north
    offsets in arcsecs from the nucleus. The TIMMI2 astrometry is
    uncertain to $\sim 5\arcsec$.}
\label{timmi}
\end{figure}

\subsection{Spectral energy distribution}\label{sed}

\begin{figure}
  \centering\includegraphics[bb=40 28 519 512,width=.8\columnwidth]{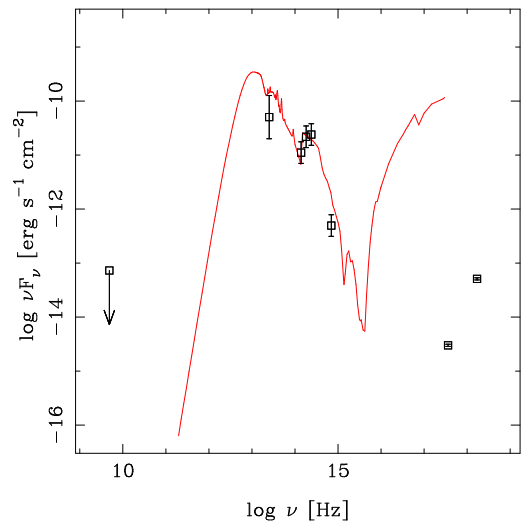}
  \caption{Spectral energy distribution of distinct nuclear source
    (AGN) in NGC\,4438. The solid line is a theoretical SED based on
    radiative transfer calculations with a central AGN as the heating
    source. See text for information about data and models. The weak
    high-energy emission from the nucleus of the galaxy suggests that
    the flows in the accretion disc are advection-dominated. }
  \label{fig:sed}
\end{figure}

NGC\,4438 is classified as a LINER~1.9 or a ``dwarf'' Seyfert 2 galaxy
on the basis of broad H$\alpha$ emission \citep{ho97,ken02}. We
constructed the NGC\,4438 SED of the non-stellar component using data
from radio ($\nu = 4.86\times 10^9$~Hz) to hard X-rays ($\nu = 1.7
\times 10^{18}$~Hz).

The near-IR data ($\log (\nu/\mathrm{Hz}) \sim 14$) were obtained from
the surface brightness decomposition presented in
Section~\ref{pho}. The AGN component photometry is given in
Table~\ref{nucphot}, and it was computed integrating the residual
image over a circular aperture of radius 4.5\arcsec\ centred on the
nucleus.

The thermal infrared ($\log (\nu/\mathrm{Hz}) \sim 13.4$) photometry
of the nuclear source was extracted over a 4.5\arcsec\ circular
aperture in the TIMMI2 NB~11.9~$\mu$m image (see Fig.~\ref{timmi}).
The result was a mid-IR flux density of $0.20 \pm 0.04$~Jy. Instead of
a compact nuclear source, Fig.~\ref{timmi} shows that the thermal
emission is spatially extended over 3.5\arcsec, which at the Virgo
distance corresponds to 270~pc. Thus, the nucleus and a circumnuclear
dusty arc dominate the mid-IR continuum emission.

The soft ($\log (\nu/\mathrm{Hz}) = 17.6$) and hard ($\log
(\nu/\mathrm{Hz}) = 18.2$) X-ray counterparts were obtained from
\textit{Chandra} data reported in Table~4 of \citet{mac04}, without
any absorption correction. The radio continumm ($\log
(\nu/\mathrm{Hz}) = 9.3$) data were obtained from \citet{hum91}. This
measurement was computed by integrating the extended emission over a
rectangle of sides $2\arcsec \times 5\arcsec$. It gives $1.5 \pm
0.3$\,mJy. This value corresponds to an upper limit since the nucleus
was not clearly detected, and it may contain emission from the NW and
SE outflows.

Fig.~\ref{fig:sed} shows the SED of NGC\,4438. The solid line is a
simple theoretical SED taken from \citet{sie04}, which is based on
three parameters: a dusty torus with radius 125\,pc, an average visual
extinction of 4 magnitudes and a compact source with bolometric
luminosity $L_{\rm bol} = 5.6 \times 10^{10}~\Lsun$, which does not
represent NGC\,4438's bolometric luminosity because of discrepancy
between with the model in the X-ray region of the spectrum. The AGN is
modelled as a power law with monochromatic luminosity
$L(\nu/\mathrm{Hz}) \propto (\nu/\mathrm{Hz})^{-0.7}$ in the
wavelength range from 10~\AA\ to 2~$\mu$m. The dust is composed of
carbon and silicate grains with radii between 300 and 2400~\AA,
graphites of radius 10~\AA\ and small and large PAH
components. Fig.~\ref{fig:sed} also shows a good match between both
near-IR and visible data with the theoretical SED. However, this model
is unable to fit the X-ray data. The sub-Eddington X-ray emission $L_X
\sim 10^{-6}~{\rm L}_{\mathrm{Edd}}$ (where ${\rm
  L}_{\mathrm{Edd}}\sim 10^{45}~\mathrm{erg~s^{-1}}$ is the Eddington
luminosity for a black hole with mass 10$^7$~\Msun) in both, the soft
and hard bands, suggests that the accretion flows are
advection-dominated \citep{nar96,ken02,mac04}.

\subsection{Outflows and energetics}\label{ss_outflow}

From UV observations, \citet{bos05} found that the star formation in
the main body of NGC\,4438 is very weak, and is mainly composed of old
stars, without signs of recent starbursts. Our estimation of the
visual extinction toward the nucleus of $A_{\rm V}\sim 4$ corresponds
to an extinction of 3.3~mag at H$\alpha$ \citep{car89}, implying an
extinction-corrected SFR of 0.08~\Msun~yr$^{-1}$ for the nucleus. On
the other hand, the SFRs inferred from the extinction-corrected
Pa$\beta$ emission is 1~\Msun~yr$^{-1}$ and corresponds to the
circumnuclear region seen as an arc in Fig.~\ref{fig:bub}a. These two
estimates correspond to upper limits since we do not know how much of
the line emission arises from {H \sc ii} regions near the nucleus.

The velocity of the northwestern outflow can be estimated from the
spectrum shown in Fig.~\ref{fig:extended}a. The width of the line
contains contributions from both random motions and bulk velocity. The
FWHM of the [Fe\,{\sc ii}] line at 3\arcsec -- 4\arcsec\ from the
nucleus, extracted from a 2\arcsec\ aperture is measured to be
$\sim$350~\kms\ while the mean [Fe\,{\sc ii}] velocity 2\arcsec\ from
the nucleus (at the bottom of distinc blob) is closer to
500~\kms\ (see Fig.~\ref{fig:extended}). In fact, the mean velocity as
a function of distance from the nucleus decreases from
$\sim$500~\kms\ at 2\arcsec\ to $\sim$350~\kms at 2.5\arcsec. This
seems consistent with an expanding bubble model since if the observed
expansion velocity at the outer edge of the bubble corresponds to a
lower line-of-sight velocity than at the middle of the bubble, due to
projection effects. Hence we adopt a speed of 500~\kms as a reasonable
value for the expansion velocity, which is a faster outflow than the
one assumed by \citet{ken02} of 300~\kms. \note{\citet{ken02}
  calculated the mass of ionised gas from the H$\alpha$ luminosity,
  which corresponds to 3.5$\times 10^4$~\Msun\ for an electron density
  of 420~cm$^{-3}$ (measured from the [S\,{\sc
      ii}]~$\lambda\lambda$6716/6731 doublet ratio).} This new
estimate of the outflow velocity increments the kinetic energy by a
factor of $\sim$3, yielding an injected kinetic energy into the halo
of the galaxy of $>10^{54}$~erg. This is a lower limit since only
ionised gas is included in the calculation and we know that there is
molecular gas also present in the NW bubble (see
Fig.~\ref{fig:bub}b). The large amount of kinetic energy carried by
the northwestern outflow along with the weak nuclear star formation,
suggests that an AGN central engine is responsible for much of the
optical and near-IR line and continuum emission, and not a compact
starburst \citep{ken02}.

\subsection{NGC\,4438 colours}\label{ss_colours}

The near-infrared colours of the nucleus of NGC\,4438 mainly have
contributions from: a bulge of late-type stars, a non-stellar nuclear
source, re-radiation from hot dust, and reddening \citep{kot92}. We
measured the colours of each surface brightness component in an
aperture of 3\arcsec, in order to compare with other works (there was
no substantial variation using an aperture of $8\arcsec$). Normal
(inactive) spiral galaxies have colours of $0.6 < J-H < 0.9$ and
$H-K_\mathrm{s} < 0.3$ \citep{kot92,fis06}. The stellar colours of
NGC\,4438, $J-H = 0.7$ and $H-K_\mathrm{s} = -0.10$, are clearly in
the region occupied by normal spiral nuclei. As expected for LINERs
and Seyfert 1 galaxies \citep[see two-colour diagrams in
][]{kot93,for92} the non-stellar colour of NGC\,4438 is located in
between the regions of AGN and inactive galaxies, shifting the colours
according to a reddened vector of $A_{\rm V}\sim 3-4$. Therefore, the
near-IR stellar colours do not show evidence for starburst, in
agreement with the UV observation of \citet{bos05}. This seems to be
inconsistent with the relatively high SFR estimated from Pa$\beta$
emission for the circumnuclear region (see Section~\ref{sec:pab}),
which would imply that this emission is not due to star formation but
to the densest gas near the nucleus being photoionised by the central
AGN.

\subsection{Black hole mass}

\citet{fer06} pointed out that a more or less a constant fraction of a
galaxy bulge mass ends up as a central massive object, either a
stellar nucleus or a supermassive black hole. The evidence discussed
in Section~\ref{sec:CO} could suggest the presence of a black hole,
although the radius of its sphere of influence $R_{\rm BH} \sim$
0.1\arcsec\ is considerably smaller than our
0.74\arcsec\ resolution. \citet{mer00} found that there is a tight
correlation between one of the fundamental properties of a galactic
bulge, its velocity dispersion $\sigma$, with the mass of its
supermassive black holes. Independently, the same correlation was
found by \citet{geb00}, in the same year. For NGC\,4438 we have
calculated the luminosity weighted velocity dispersion to be $\sigma =
142 \pm 8$~\kms\ in the central 3 arcsec over the minor-axis, by using
the kinematical data (see Fig.~\ref{fig:vel}) and by assuming an axial
ratio $q=0.57$ (see Table~\ref{bestfit}). The $\mathcal{M}_{\rm
  bh}$-$\sigma$ relation using this value yields a mass for the
central black hole of $\log(\mathcal{M}_{\rm bh}/\Msun) = 7.5 \pm
1.7$.

Recently, \citet{gra07} suggested that there is a fundamental
correlation between the S\'ersic index $n$, which is a measure of the
concentration within the bulge, and the black hole mass. The physical
interpretation is that steeper and more concentrated bulges (larger
$n$) host more massive black holes. They found a
$\mathcal{M}_\mathrm{bh}$-$n$ log-quadratic relation using updated
black hole masses and power-law indexes. This relation was computed
from a sample of 27 galaxies with black hole mass determination from
the correlation between stellar velocity dispersion and virial bulge
mass. The S\'ersic indexes in \citet{gra07} were determined by
performing a surface brightness decomposition similar to the one used
in this work. The $\mathcal{M}_\mathrm{bh}$-$n$ relation is as tight
as the well known correlation between the stellar velocity dispersion
and the black hole mass \citep{gra07}.

We found a bulge well-represented by a S\'ersic index $n=1.7$. This
index implies a black hole with mass $\log(\mathcal{M}_{\rm bh}/\Msun)
= 6.70\pm 0.45$ for NGC\,4438 (for the estimation in each broadband
see column (16) of Table~\ref{bestfit}). The errors in the black hole
mass were estimated using the intrinsic uncertainties in the
\citet{gra07} relation. We did not estimate a black hole mass from the
bulge parameters inferred from the $B$-band image, because the bulge
is highly obscured by dust. Another estimator of $\mathcal{M}_{\rm
  bh}$ is the near-IR luminosity \citep{mar03}, which provides a
$\mathcal{M}_{\rm bh}$-$L_{\rm bulge}$ relation tighter and less
sensitive to extinction than those in the optical. Column (15) of
Table~\ref{bestfit} lists the computed black hole masses using the
near-IR luminosity. It can be seen that these methods yield results in
reasonable good agreement.


\section{Conclusions}

In this paper we have presented a study of the nuclear source and
central environment of the galaxy NGC\,4438, based on the results of
near-infrared ISAAC VLT/ANTU imaging and spectroscopy. The main
results of this work are listed below.
\begin{enumerate}
\item We have found extended [Fe\,{\sc ii}] emission coincident with
  the NW bubble seen in radio continuum, X-rays and optical
  emission. The morphological correlation between [Fe\,{\sc ii}] and
  previous radio observations, along with the absence of Pa$\beta$ in
  our spectra of the NW bubble, suggest that shocks (perhaps driven by
  a radio jet) may be an important source of excitation of [Fe\,{\sc
      ii}] emission.

\item Based on our newly (upwardly) revised estimated of the expansion
  velocity of the bubble, 500~\kms, and the nuclear star formation
  rate estimated from its emission in H$\alpha$, which could be up to
  0.08~\Msun~yr$^{-1}$, corrected by the nuclear extinction, we have
  addressed the question whether the outflow has an AGN or a starburst
  origin. The large kinetic energy associated with the outflowing gas,
  $>$10$^{54}$~erg, along with the weak star formation suggest that an
  AGN is more likely to power the outflow.

\item The H$_2$ emission map shows strong molecular hydrogen emission
  around the nucleus which might indicate the presence of a molecular
  torus. 

\item Our position-velocity maps showed that the gas emitting H$_2$ is
  kinematically distinct from the gas radiating [Fe\,{\sc ii}]
  emission. Moreover, the nuclear [Fe\,{\sc ii}] line (FWHM $\sim$
  415~\kms) is broader than the nuclear H$_2$ emission (FWHM $\sim$
  270~\kms), implying that the [Fe\,{\sc ii}] gas is kinematically
  more disturbed than the H$_2$ gas. Furthermore, the molecular gas
  seems to be more restricted to the galactic plane, while [Fe\,{\sc
      ii}] extends to higher latitudes from the plane of the
  galaxy. These results suggest that X-ray heating from a central AGN
  may be responsible for the excitation of H$_2$.
  
\item We have applied a two-dimensional surface brightness
  decomposition to the central 30 arcsec of NGC\,4438. The best-fitted
  model consists of three components: a S\'ersic bulge with power-law
  index $n\sim 1.7$ and $R_e\sim 3$\arcsec, an exponential disc with
  $R_h\sim 13$\arcsec, and a compact nuclear source, resolved in
  \textit{HST} with $R_e\sim 0.2$\arcsec. The model was evaluated in a
  generalised elliptical surface with discy isophotes and an
  inclination angle of $69^{\circ}$.

\item We have constructed the spectral energy distribution of the
  nucleus of NGC\,4438. It is in agreement with a theoretical SED of
  an AGN obscured by 4~magnitudes, although the X-rays are
  underluminous by 6 orders of magnitude (compared to the Eddington
  luminosity for a 10$^7$~\Msun\ black hole), which can be explained
  on the basis of advection-dominated accretion flows.

\item We have derived CO-band stellar line-of-sight kinematics along
  the minor axis of the galaxy. The line-of-sight velocity dispersion
  showed a strong peak on the centre of the galaxy of $177\pm8$~\kms.

\item A black hole mass of $\log(\mathcal{M}_{\mathrm{bh}}/\Msun) \sim
  7$ for the nuclear point source could be inferred from the bulge
  luminosity, the central velocity dispersion and the S\'ersic index
  $n$.

\item We have carried out near-IR photometry of the nucleus, finding a
  point source unresolved by ISAAC at 0.8\arcsec\ resolution with
  \mbox{$M_K=-18.7$} and $J-K = 0.69$. This detection is further
  evidence that the central source is associated with an AGN rather
  than a starburst engine.

\item We report the brightnesses and colours of the various components
  in the central region. The dereddened stellar colors (bulge and
  disk) in the nuclear region are typical of an inactive spiral
  galaxy, and show no evidence for a recent starburst. This is further
  evidence that significant star formation was not triggered in the
  recent collision with M86.

\end{enumerate}

\section*{Acknowledgements}
We thank the referee for valuable comments which helped to improve
this paper. We are very grateful to Lowell Tacconi-Garman for
providing calibration data and useful comments on the reduction of the
ISAAC data. We thank Andrew Baker and Eduardo Hardy for valuable input
to the proposal. We would like to thank Katherine Blundell for helpful
comments on the manuscript. S.P. acknowledge generous support from
STFC and CONICYT.  S.C. acknowledges support from FONDECYT grant
1060827, and from the Chilean Center for Astrophysics FONDAP
15010003. Thanks also to Andr\'es Jordan for the \texttt{PDL::Minuit}
package. This research also used data products from the Canadian
Astronomy Data Centre operated by the National Research Council of
Canada with the support of the Canadian Space Agency.

\appendix

\section[]{Data Reduction}\label{dr}

In this Appendix we outline the procedures used in the data reduction
process, which included bias subtraction, flat fielding, correction of
detector defects, cosmic rays removal, and overlapping of the dithered
frames. The detector has a jump between the two halves of the array,
caused by imperfect removing of the zero level offset due to
variations in the bias level \citep[][see below]{ami01}.

\noindent\textit{Flat fielding and bad pixels correction.--} The
twilight flat field images provided by the observatory were sometimes
affected by spurious large scale gradients (specially in the
$K_\mathrm{s}$-band). This could be caused by bad ambient conditions,
making non-linear effects important (L. Tacconi, private
communication). We thus divided the flat field by a smooth
approximation of the large scale variations. The smooth flat field was
generated by taking the median of the detector image along the
$x$-axis, excluding the outer $\sim100$ columns, which generated a
vector column representative of the large scale variation.

\noindent\textit{Sky and bias residual subtraction.--} This is the
most important step in the reduction. The sky was obtained immediately
after the last object image, pointing the telescope towards a blank
region of the sky. The subtraction of the dark current was one of the
most important problems to solve, because the dark current is known to
be unstable in ISAAC \citep{ami01}. The bias level of the detector is
a function of the detector integration time and its illumination. The
bias level also varies in time, and is more pronounced where the
readouts start (i.e., rows 1, 2, and rows 513, 514). Due to this
variation we did not apply the dark subtraction, which yielded images
containing noticeable bias residuals. We removed these bias artifacts
by performing a linear extrapolation along the array. The procedure
was:
\begin{enumerate}
\item[1.] to isolate regions devoid of extended emission,
\item[2.] then, to smooth the regions with a median filter and
  collapse them along the $x$-axis to obtain single bias residual
  columns,
\item[3.] finally, perform a linear interpolation between the
  collapsed columns to the entire detector array, generating a smooth
  bias image.
\end{enumerate}
\noindent\textit{Dithering.--} The shifts in the dithered frames were
obtained by maximizing the covariance between the images as a function
of astrometric offset (at a sub pixel level).\newline

\noindent\textit{Photometric calibration}

We acquired a dedicated set of relatively faint standards from the
UKIRT telescope system \citep{ukirt}, specifically the stars FS6, FS20
($J$, $H$ filters) and FS132 ($K_\mathrm{s}$ filter). The three
standard star observations share a similar dithering pattern as the
object observations, and they were reduced following the procedure
explained above. The calibration was based on aperture photometry
(aperture radius 4.4\arcsec). Table~\ref{tab:calib} summarises the
computed zero points. We calibrated the narrowband filters using the
corresponding broadband calibration \citep[as recommended in the ISAAC
  manual,][]{ami01}. No corrections for atmospheric extinction were
required because the data were obtained at similar airmass
($\sim1.4$).

Foreground Galactic extinction obtained from the \citet{dust} maps of
dust emission, amounts to $E(B-V) = 0.028$. Table~\ref{tab:calib}
summarises the magnitudes of extinction for each bandpass, computed
using $R_{\mathrm{V}} = 3.1$ and the extinction laws of \citet{car89}.

Photometric uncertainties were estimated as $\sigma_t^2 =
\sigma^2_{\mathrm{rms}} + \sigma^2_{\mathrm{calib}}$. The first terms
is the rms noise and the second term takes into account the
calibration accuracy (10 per cent).

We verified the measured fluxes in our calibrated images using
2MASS. The result was a $2 - 4$ per cent difference between 2MASS
photometry and our photometry of the standard stars named above, and
differences of 12, 14 and 10 per cent for $JHK_\mathrm{s}$,
respectively for NGC\,4438. This could be caused by an overestimation
of the sky level in our reduction procedure.

\section{Surface brightness profile model}\label{sbm}

\begin{figure}
  \centering\includegraphics[bb=66 53 409 444,width=0.8\columnwidth]{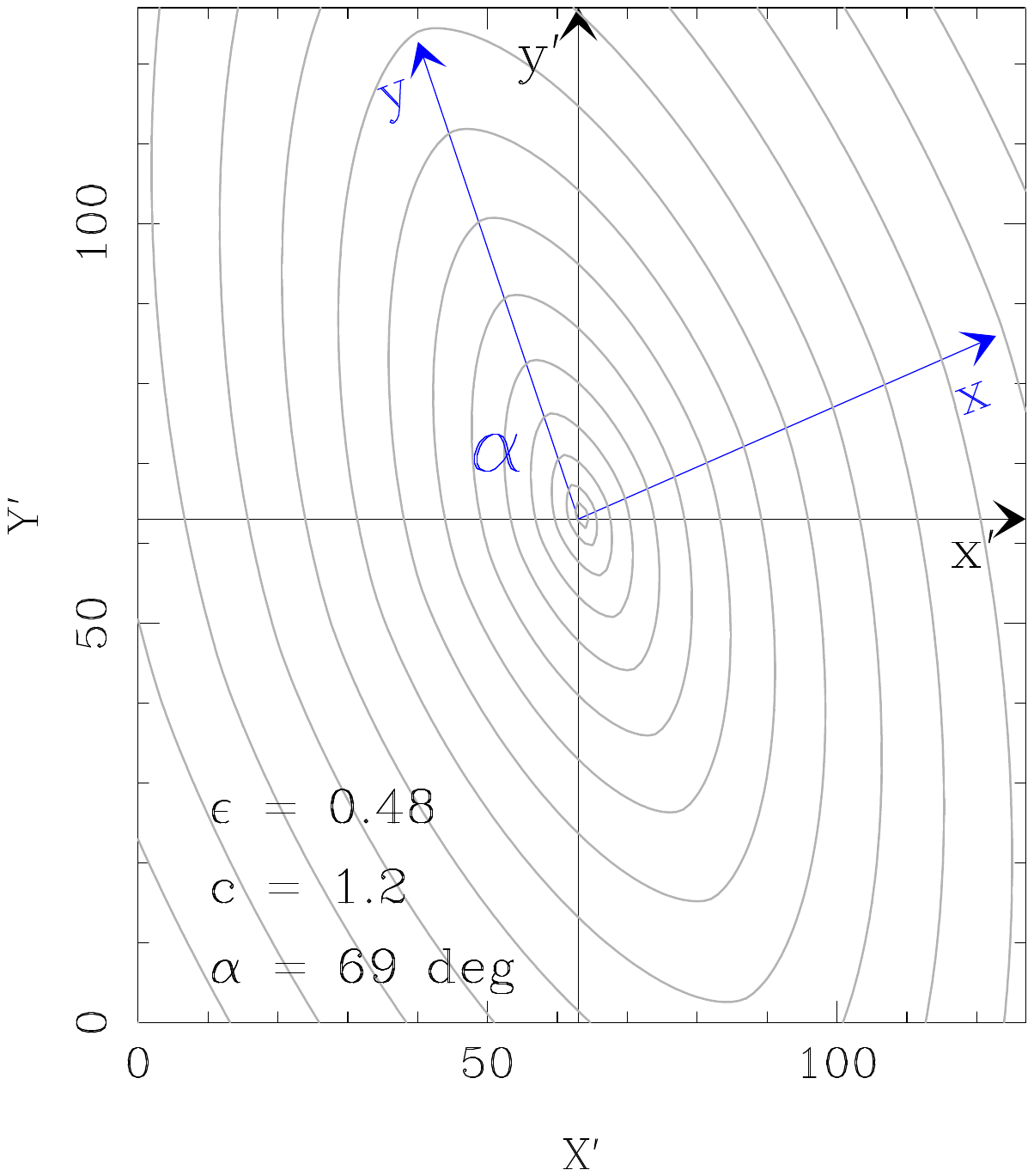}
  \caption{Best-fitted isophotes model to NGC\,4438 with constant
    ellipticity $\epsilon=0.48$, angle $\alpha=69^{\circ}$ and
    $c=1.2$. This figure also shows the convention used for the pixel
    coordinates $\mathbf{X'}$, $\mathbf{X}$, and the angle $\alpha$.}
    \label{xy}
\end{figure}

This Appendix gives a detailed description of the surface brightness
model used to fit the NGC\,4438 light describution. Our conventions
and coordinate systems are shown in Fig.~\ref{xy}. In order to obtain
an accurate fit we constructed an elliptical polar-grid, with the
coordinate axis defined by \citet{atha90}, as follows.

\begin{equation}\label{rad}
  r = \left( \left|x\right|^{2/c} + \left|y/q
    \right|^{2/c}\right)^{c/2},
\end{equation}

\noindent where $q$ is related to the ellipticity by $e = 1-q$. $c$ is
the parameter which allows us to model the generalised elliptical
shape: $c<1$ gives box-like ellipses (boxiness), while for $c>1$ we
obtain shapes approaching to diamonds (diskiness). In general,
galaxies are inclined at an angle $\alpha$, so we need to rotate our
$x$- and $y$-axis using a simple transform; $\mathbf{X} = R(\alpha)
\mathbf{X'}$, where $ R(\alpha)$ is the rotation matrix (see
Fig.~\ref{xy}). The angle $\alpha$ is defined with respect to the
image pixel coordinate system, increasing counterclockwise.

The \citet{ser68} light profile is very useful for modelling
elliptical galaxies and steep spiral bulges. It has the form:

\begin{equation}
  I_b(r) = I_e\,\exp{\left\{-b_n \left[
      \left(\frac{r}{R_e}\right)^{1/n}-1\right]\right\}},
  \label{sersic}
\end{equation}

\noindent where $r$ is the generalised elliptical radius (defined in
Equation~\ref{rad}), $I_e$ is the intensity at the effective radius
$R_e$ and $b_n$ is defined such that $\Gamma (2n) = 2\,\gamma(2n,b_n)$
where $\Gamma$ and $\gamma$ are the complete and incomplete gamma
functions, respectively. Analytical expressions which approximate the
value of $b_n$ give $b_n = 2 \, n - 0.33$, for $0.5 < n < 10$. Half of
the total luminosity predicted by the profile comes from $r<R_e$. The
power-law index $n$ describes the shape of the light-profile. To
calculate the total flux of the S\'ersic profile, we integrated over a
generalised elliptical radius $r$. It gives:

\begin{equation}
  F_{\nu}(r) = \pi \, I_e \, e^{b_n}\,R_e^2\, \frac{2n}{b_n^{2n}}\,
  \gamma (2n,x)\, \mathcal{I}(q,c),
  \label{Ftot}
\end{equation}

\noindent where $x=b_n (r/R_e)^{1/n}$. The function $\mathcal{I}(q,c)$
is given by

\begin{equation}
  \mathcal{I}(q,c) = \frac{2 q c}{\pi}\, \int_0^{\pi/2} \left( \sin
    \phi \,\cos \phi \right)^{c-1}\, d\phi.
  \label{int}
\end{equation}

This expression accounts for the generalised-elliptical shape. When
$c$ is greater than 2 (`disky' isophotes) the integral can be written
as a beta function of the parameter $c$ \citep[see Equation 8 in
][]{pen02}.

Since NGC\,4438 is a spiral galaxy, an exponential law is needed to
fit the galactic disc. It has the form $I_d(r) = I_0\,\exp{(-r /
  R_h)}$, where the parameter $R_h$ is called the disc scale-length,
$I_0$ is the peak of luminosity, and $r$ is the elliptical radius
(Equation~\ref{rad}).

A nuclear source can be added at the centre of the galaxy,
representing the AGN component, and it may be either a delta function
(for a point source) or a S\'ersic component (for resolved sources).
Finally, the model takes the form:

\begin{equation}\label{model}
  I(r) = \left (I_\mathrm{nuc} (r) + I_b (r) + I_d (r) \right )
  \otimes\, \mathrm{PSF},
\end{equation}

\noindent where $I(r)$ is the surface brightness at a radius $r$
(defined in Equation~\ref{rad}), $I_\mathrm{nuc}$ represents the
nuclear source, $I_b$ and $I_d$ represent the bulge and disc
components, respectively. The symbol $\otimes$ is the convolution
operator.

The set of free parameters are: the nuclear source centroid
($x_\mathrm{cen},y_\mathrm{cen}$), those related to the elliptical
geometry: $\alpha$, $q$, $i$ and $c$; the S\'ersic profile power-law
index $n$, the effective radius $R_e$, the effective intensity $I_e$;
the exponential disc scale-length $R_h$ and central luminosity $I_0$;
the nuclear source parameters: $n'$, $R_e'$ and $I_e'$ (S\'ersic case)
or the central flux density (delta function case).

We found the best fitting solution by performing a $\chi^2$
minimisation. The main steps of the two-dimensional fitting are
roughly summarised as follows:
\begin{enumerate}
\item[1.] Select a sub-image, centred on the nucleus. Correct by
  extinction.
\item[2.] Generate a model image based on the initial (or new)
  conditions.
\item[3.] Account for the telescope and atmospheric seeing by
  convolving the model with the characteristic PSF.
\item[4.] Evaluate $\chi^2$.
\item[5.] Iterate from step 2 until convergence is reached.
\item[6.] Generate output images: residual image, original galaxy
  image and model.
\end{enumerate}

\bsp

\label{lastpage}

\end{document}